\documentclass{cidr-2019-arxiv}

\pdfoutput=1

\usepackage{xcolor}
\usepackage{graphicx}
\usepackage{subcaption}
\usepackage{epstopdf}
\usepackage{wrapfig}
\usepackage{hyperref}
\usepackage{array}
\usepackage{nicefrac}
\usepackage{fixme}
\usepackage{amssymb}
\usepackage{tikz}
\usepackage{multirow}
\usepackage{booktabs} 

\usepackage{wrapfig}
\usepackage{colortbl}
\usepackage{algorithmic}

\usepackage{cleveref}
\crefformat{section}{(\S#2#1#3)}

\usepackage{epigraph}

\usepackage{mathtools}

\usepackage{tikz}

\usepackage{subfloat}

\usepackage{scalefnt}
\usepackage{bm}
\usepackage{amsmath}

\usepackage{float}
\usepackage{dblfloatfix}

\usepackage{afterpage}

\usepackage{lscape}

\makeatletter
\newcommand\notsotiny{\@setfontsize\notsotiny\@vipt\@viipt}
\makeatother

\newcommand\blue[1]{\textcolor{blue}{#1}}

\usetikzlibrary{matrix}
\usetikzlibrary{positioning}
\usetikzlibrary{fit}
\usetikzlibrary{calc}

\usepackage[vlined,linesnumbered]{algorithm2e}

\SetCommentSty{mycommfont}
\SetAlFnt{\sffamily}

\SetAlCapFnt{\normalfont\sffamily\small}

\makeatletter
\renewcommand{\algocf@Vline}[1]{
  \strut\par\nointerlineskip
  \algocf@push{\skiprule}
  \hbox{\bgroup\color{cyan}\vrule\egroup%
    \vtop{\algocf@push{\skiptext}
      \vtop{\algocf@addskiptotal #1}\bgroup\color{cyan}\Hlne\egroup}}\vskip\skiphlne
  \algocf@pop{\skiprule}
  \nointerlineskip}
\renewcommand{\algocf@Vsline}[1]{
  \strut\par\nointerlineskip
  \algocf@bblockcode%
  \algocf@push{\skiprule}
  \hbox{\bgroup\color{cyan}\vrule\egroup
    \vtop{\algocf@push{\skiptext}
      \vtop{\algocf@addskiptotal #1}}}
  \algocf@pop{\skiprule}
  \algocf@eblockcode%
}
\makeatother
\SetKwProg{Fn}{Function}{}{}

\fxsetup{
    status=draft,
    author={Todo},
    layout=inline,
    theme=color,
    silent=true,
    }
\definecolor{fxnote}{rgb}{0.8000,0.0000,0.0000}

\newcommand{\showDOI}[1]{\unskip}
\newcommand{\showURL}[1]{\unskip}
\newcolumntype{x}[1]{>{\raggedright}p{#1}} 

\graphicspath{{graphics/}}

\begin{document}

\title{Learning Key-Value Store Design}

\author{Stratos Idreos, Niv Dayan, Wilson Qin, Mali Akmanalp, Sophie Hilgard, Andrew Ross, \\ James Lennon, Varun Jain, Harshita Gupta, David Li, Zichen Zhu \\\\
Harvard University
}

\maketitle

\begin{abstract}
We introduce the concept of design continuums for the data layout of key-value stores. A design continuum unifies major distinct data structure designs under the same model. The critical insight and potential long-term impact is that such unifying models 1)~render what we consider up to now as fundamentally different data structures to be seen as ``views'' of the very same overall design space, and 2)~allow ``seeing'' new data structure designs with performance properties that are not feasible by existing designs. The core intuition behind the construction of design continuums is that all data structures arise from the very same set of fundamental design principles, i.e., a small set of data layout design concepts out of which we can synthesize any design that exists in the literature as well as new ones. We show how to construct, evaluate, and expand, design continuums and we also present the first continuum that unifies major data structure designs, i.e., B$^{+}$tree, B$^{\epsilon}$tree, LSM-tree, and LSH-table.

The practical benefit of a design continuum is that it creates a fast inference engine for the design of data structures. For example, we can predict near instantly how a specific design change in the underlying storage of a data system would affect performance, or reversely what would be the optimal data structure (from a given set of designs) given workload characteristics and a memory budget. In turn, these properties allow us to envision a new class of self-designing key-value stores with a substantially improved ability to adapt to workload and hardware changes by transitioning between drastically different data structure designs to assume a diverse set of performance properties at will.   
\end{abstract}

\section{A Vast Design Space}
\label{sec:intro}

\textbf{Key-value stores are everywhere}, providing the storage backbone for an ever-growing number of diverse applications. The scenarios range from graph processing in social media \cite{Armstrong2013, Bu2014}, to event log processing in cybersecurity \cite{Cao2013}, application data caching \cite{Memcached}, NoSQL stores \cite{Redis}, flash translation layer design \cite{Dayan2016}, time-series management \cite{Kondylakis2018, Kondylakis2019}, and online transaction processing \cite{Dong2017}. In addition, key-value stores increasingly have become an attractive solution as embedded systems in complex data-intensive applications, machine learning pipelines, and larger systems that support more complex data models. For example, key-value stores are utilized in SQL systems, e.g., FoundationDB \cite{AppleFoundationDB} is a core part of Snowflake \cite{Snowflake2016}, while MyRocks integrates RockDB in MySQL as its backend storage.

\begin{figure}[t]
\center
    \includegraphics[width=.4\textwidth]{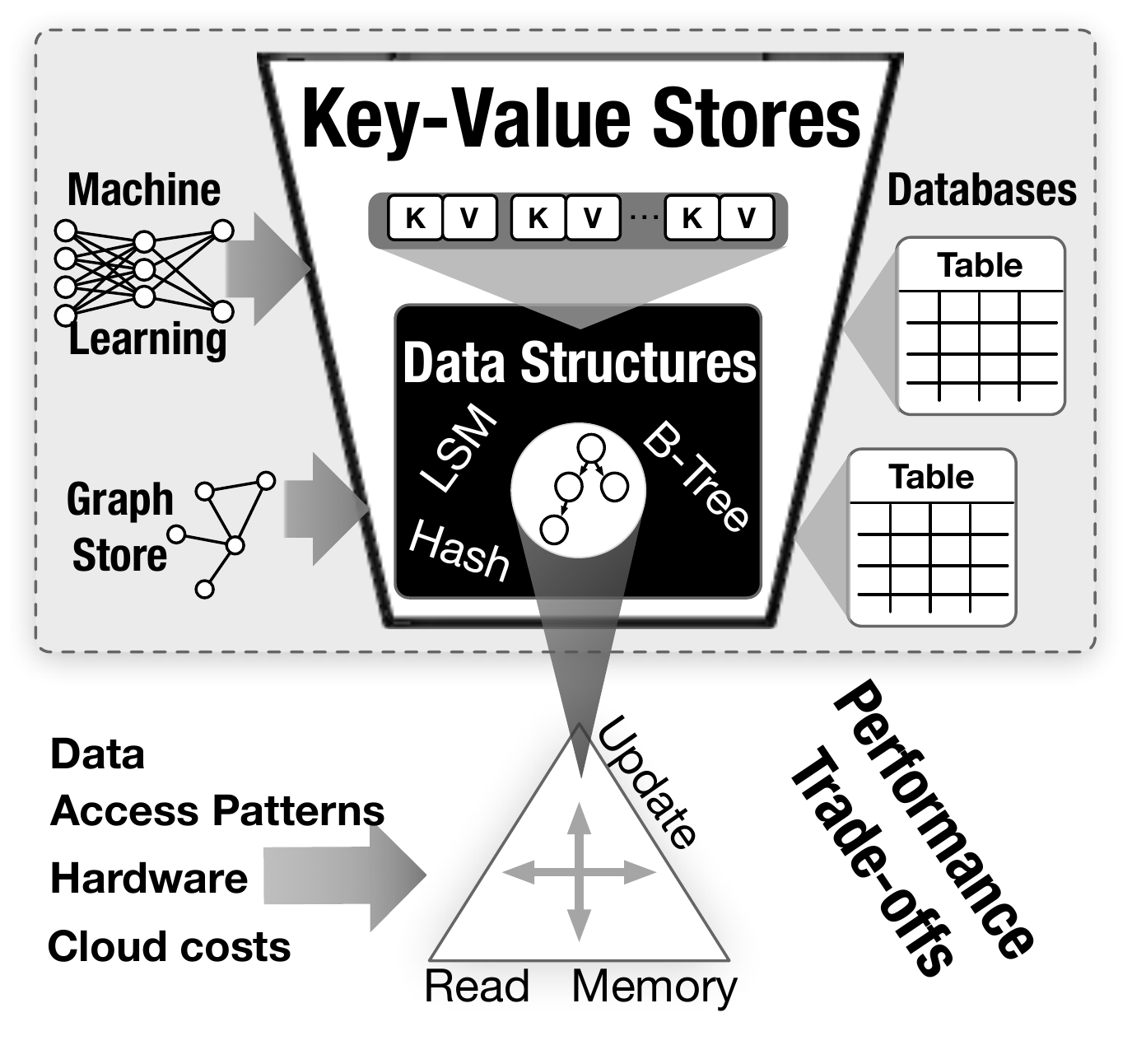}
    \vspace{-2em}
    \caption{From performance trade-offs to data structures, key-value stores and rich applications.}
    \label{f:hierarchy1}
\end{figure}

\textbf{There is no Perfect Design}. As shown in Figure \ref{f:hierarchy1}, at its core a key-value store implements a data structure that stores key-value pairs. Each data structure design achieves a specific balance regarding the fundamental trade-offs of read, update, and memory amplification \cite{Athanassoulis2016}. For example, read amplification is defined as ``how much more data do we have to read for every key we are looking for on top of accessing this particular key''. There exists no perfect data structure that minimizes all three performance trade-offs \cite{Athanassoulis2016, Idreos2018a}. For example, if we add a log to support efficient out of place writes, we sacrifice memory/space cost as we may have duplicate entries, and read cost as future queries have to search both the core data structure and the log. 

In turn, this means that there exists no perfect key-value store that covers diverse performance requirements. Every design is a compromise. But then how do we know which design is best for an application, e.g., for specific data, access patterns, hardware used, or even a maximum financial budget on the cloud? And do we have enough designs and systems to cover the needs of emerging and ever-changing data-driven applications? This is the problem we study in this paper and envision a research path that makes it easier to create custom data structure designs that match the needs of new applications, hardware, and cloud pricing schemes.

\textbf{The Big Three.} As of 2018, there are three predominant data structure designs for key-value stores to organize data. To give an idea of the diverse design goals and performance balances they provide, we go briefly through their core design characteristics. The first one is the \textbf{B$^{+}$tree} \cite{Bayer1972}. The prototypical B$^{+}$tree design consists of a leaf level of independent nodes with sorted key-value pairs (typically multiple storage blocks each) and an index (logarithmic at the number of leaf nodes) which consists of nodes of fractional cascading fence pointers with a large fanout. 
For example, B$^{+}$tree is the backbone design of the BerkeleyDB key-value store \cite{Olson1999}, now owned by Oracle, and the backbone of the WiredTiger key-value store \cite{WiredTiger}, now used as the primary storage engine in MongoDB \cite{MongoDB}. FoundationDB \cite{AppleFoundationDB} also relies on a B$^{+}$tree. Overall, B$^{+}$tree achieves a good balance between read and write performance with a reasonable memory overhead that is primarily due to its fill factor in each node (typically 50\%) and the auxiliary internal index nodes. 

In the early 2000s, a new wave of applications emerged requiring faster writes, while still giving good read performance. At the same time, the advent of flash-based SSDs has made write I/Os 1-2 orders of magnitude costlier than read I/Os \cite{Agrawal2008}. These workload and hardware trends led to two data structure design decisions for key-value stores: 1)~buffering new data in memory, batching writes in secondary storage, and 2) avoiding global order maintenance. This class of designs was pioneered by the \textbf{Log-Structured Merge Tree} (LSM-tree) \cite{ONeil1996} which partitions data temporally in a series of increasingly larger levels. Each key-value entry enters at the very top level (the in-memory write buffer) and is sort-merged at lower levels as more data arrives. In-memory structures such as Bloom filters, fence pointers and Tries help filter queries to avoid disk I/O \cite{Dayan2017, Zhang2018a}. 
This design has been adopted in numerous industrial settings including LevelDB \cite{GoogleLevelDB} and BigTable \cite{Chang2006} at Google, RocksDB \cite{FacebookRocksDB} at Facebook, Cassandra \cite{Lakshman2010}, HBase \cite{HBase2013} and Accumulo \cite{ApacheAccumulo} at Apache, Voldemort \cite{LinkedInVoldemort} at LinkedIn, Dynamo \cite{DeCandia2007} at Amazon, WiredTiger \cite{WiredTiger} at MongoDB, and bLSM \cite{Sears2012} and cLSM \cite{Golan-Gueta2015} at Yahoo, and more designs in research such as SlimDB \cite{Ren2017}, WiscKey \cite{Lu2016}, Monkey \cite{Dayan2017, Dayan2018a}, Dostoevsky \cite{Dayan2018}, and LSM-bush \cite{Dayan2019}. 
Relational databases such as MySQL and SQLite4 support this design too by mapping primary keys to rows as values. 
Overall, LSM-tree-based designs achieve better writes than B$^{+}$tree-based designs but they typically give up some read performance (e.g., for short-range queries) given that we have to look for data through multiple levels, and they also give up some memory amplification to hold enough in-memory filters to support efficient point queries. Crucial design knobs, such as fill factor for B$^{+}$tree and size ratio for LSM-tree, define the space amplification relationship among the two designs. 

More recently, a third design emerged for applications that require even faster ingestion rates. The primary data structure design decision was to drop order maintenance. Data accumulates in an in-memory write buffer. Once full, it is pushed to secondary storage as yet another node of an ever-growing single level log. An in-memory index, e.g., a hash table, allows locating any key-value pair easily while the log is periodically merged to enforce an upper bound on the number of obsolete entries. This \textbf{Log-Structured Hash-table} (LSH-table) is employed by BitCask \cite{Sheehy2010} at Riak, Sparkey \cite{SpotifySparkey} at Spotify, FASTER \cite{Chandramouli2018} at Microsoft, and many more systems in research \cite{Rumble2014, Lim2014, Ahn2016}. Overall, LSH-table achieves excellent write performance, but it sacrifices read performance (for range queries), while the memory footprint is also typically higher since now all keys need to be indexed in-memory to minimize I/O needs per key.


\textbf{The Practical Problem.} While key-value stores continue to be adopted by an ever-growing set of applications, each application has to choose among the existing designs which may or may not be close to the ideal performance that could be achieved for the specific characteristics of the application. This is a problem for several increasingly pressing reasons. First, new applications appear many of which introduce new workload patterns that were not typical before. Second, existing applications keep redefining their services and features which affects their workload patterns directly and in many cases renders the existing underlying storage decisions sub-optimal or even bad. Third, hardware keeps changing which affects the CPU/bandwidth/latency balance. Across all those cases, achieving maximum performance requires low-level storage design changes. This boils down to the one size does not fit all problem, which holds for overall system design \cite{Stonebraker2005a} and for the storage layer \cite{Athanassoulis2016}. 

Especially in today's cloud-based world, even designs \\slightly sub-optimal by 1\% translate to a massive loss in energy utilization and thus costs \cite{Kossman2018}, even if the performance difference is not directly felt by the end users. This implies two trends. First, getting as close to the optimal design is critical. Second, the way a data structure design translates to cost needs to be embedded in the design process as it is not necessarily about achieving maximum query throughput, but typically a more holistic view of the design is needed, including the memory footprint. 
Besides, the cost policy varies from one cloud provider to the next, and even for the same provider it may vary over time. 
For example, Amazon AWS charges based on CPU and memory for computation resources, and based on volume size, reserved throughput, and I/O performed for networked storage. 
Google Cloud Platform, while charging similarly for computation, only charges based on volume size for networked storage.
This implies that the optimal data structure 1) is different for different cloud providers where the key-value store is expected to run, and 2) can vary over time for the same cloud provider even if the application itself and underlying hardware stay the same. 

\textbf{The Research Challenge.}
The long-term challenge is whether we can easily or even automatically find the optimal storage design for a given problem. This has been recognized as an open problem since the early days of computer science. In his seminal 1978 paper, Robert Tarjan includes this problem in his list of the five major challenges for the future (which also included $P$ Vs $NP$) \cite{Tarjan1978}: \textit{``Is there a calculus of data structures by which one can choose the appropriate data representation and techniques for a given problem?''}. We propose that a significant step toward a solution includes dealing with the following \textbf{two challenges:} 
    \vspace{.1em}

\textbf{1)}~Can we know all possible data structure designs? 
    \vspace{.5em}
    	
\textbf{2)}~Can we compute the performance of any design? 
    \vspace{1em}

\begin{figure}[t]
    \hspace{-1em}
    \includegraphics[width=.49\textwidth]{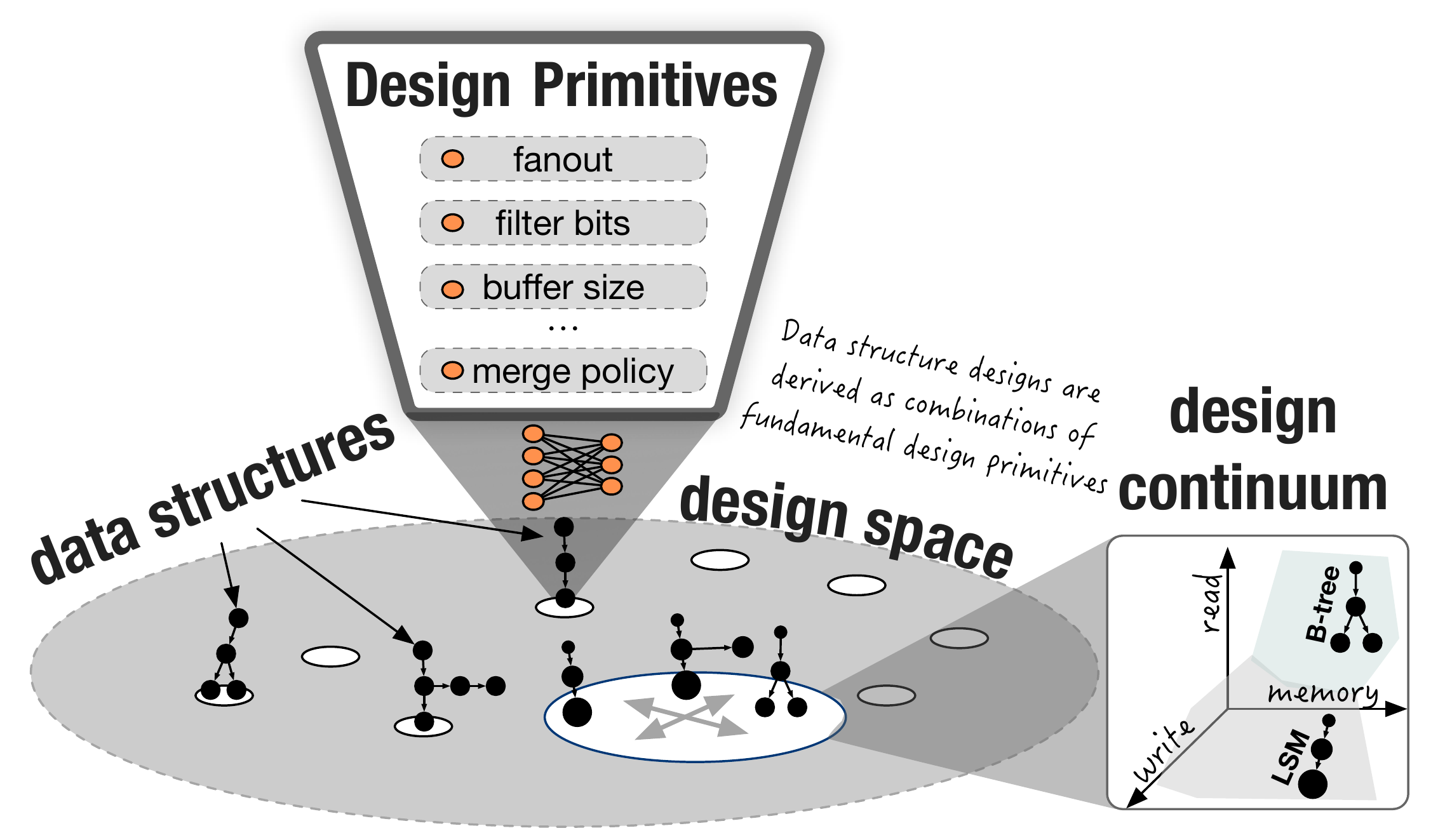}
    \vspace{-2em}
    \caption{From data layout design principles to the design space of possible data structures, where design continuums can be observed to help navigate performance trade-offs across diverse data structure designs.}
    \label{f:hierarchy}
\end{figure}

\textbf{Toward an Answer to Challenge 1.}
We made a step toward the first challenge by introducing the \textbf{design space} of data structures supporting the key-value model \cite{Idreos2018}. The design space is defined by all designs that can be described as combinations and tunings of the ``first principles of data layout design''. A first principle is a fundamental design concept that cannot be broken into additional concepts, e.g., fence pointers, links, and temporal partitioning. The intuition is that, over the past several decades, researchers have invented numerous fundamental design concepts such that a plethora of new valid designs with interesting properties can be synthesized out of those \cite{Idreos2018}. 

As an analogy consider the periodic table of elements in~chemistry; it sketched the design space of existing elements based on their fundamental components, and allowed researchers to predict the existence of unknown, at the time, elements and their properties, purely by the structure of the design space. In the same way, we created the \textbf{periodic table of data structures} \cite{Idreos2018a} which describes more data structure designs than stars on the sky and can be used as a design and new data structure discovery guide. 

Naturally, a design space does not necessarily describe ``all possible data structures''; a new design concept may be invented and cause an exponential increase in the number of possible designs. However, after 50 years of computer science research, the chances of inventing a fundamentally new design concept have decreased exponentially; many exciting innovations, in fact, come from utilizing a design concept that, while known, it was not explored in a given context before and thus it revolutionizes how to think about a problem. Using Bloom filters as a way to filter accesses in storage and remote machines, scheduling indexing construction actions lazily \cite{Idreos2007}, using ML models to guide data access \cite{Kraska2018}, storage \cite{Kersten2017} and other system components \cite{Kraska2019}, can all be thought of as such examples.
 Design spaces that cover large fundamental sets of concepts can help accelerate progress with figuring out new promising directions, and when new concepts are invented they can help with figuring out the new possible derivative designs. 

\textbf{Toward an Answer to Challenge 2.}
The next piece of the puzzle is to investigate if we can make it easy to compute the performance properties of any given data structure design. 
With the Data Calculator we introduced the idea of \textbf{learned cost models} \cite{Idreos2018} which allow learning the costs of fundamental access patterns (random access, scan, sorted search) out of which we can synthesize the costs of complex algorithms for a given data structure design. These costs can, in turn, be used by machine learning algorithms that iterate over machine generated data structure specifications to label designs, and to compute rewards, deciding which design specification to try out next. Early results using genetic algorithms show the strong potential of such approaches \cite{Idreos2017}. However, there is still an essential missing link; given the fact that the design space size is exponential in the number of design principles (and that it will likely only expand over time), such solutions cannot find optimal designs in feasible time, at least not with any guarantee, leaving valuable performance behind \cite{Kossman2018}. This is the new problem we attack in this paper: Can we develop fast search algorithms that automatically or interactively help researchers and engineers find a close to optimal data structure design for a key-value store given a target workload and hardware environment? 

\textbf{Design Continuums.}
Like when designing any algorithm, the key ingredient is to induce domain-specific knowledge. Our insight is that there exist ``design continuums'' embedded in the design space of data structures. An intuitive way to think of design continuums is as a performance hyperplane that connects a specific subset of data structures designs. Design continuums are effectively a projection of the design space, a ``pocket'' of designs where we can identify unifying properties among its members. Figure \ref{f:hierarchy} gives an abstract view of this intuition; it depicts the design space of data structures where numerous possible designs can be identified, each one being derived as a combination of a small set of fundamental design primitives and performance continuums can be identified for subsets of those structures.  

\begin{enumerate}
\item We introduce design continuums as subspaces of the design space which connect more than one design. A design continuum has the crucial property that it creates a continuous performance tradeoff for fundamental performance metrics such as updates, inserts, point reads, long-range and short-range scans, etc.
\item We show how to construct continuums using few design knobs. For every metric it is possible to produce a closed-form formula to quickly compute the optimal design. Thus, design continuums enable us to \textbf{know} the best key-value store design for a given workload and hardware. 
\item We present a design continuum that connects major classes of modern key-value stores including LSM-tree, B$^{\epsilon}$tree, and B$^{+}$tree.
\item We show that for certain design decisions key-value stores should still rely on \textbf{learning} as it is hard (perhaps even impossible) to capture them in a continuum. 
\item We present the vision of self-designing key-value stores, which morph across designs that are now considered as fundamentally different.

\end{enumerate}

\textbf{Inspiration.}
Our work is inspired by numerous efforts that also use first principles and clean abstractions to understand a complex design space. John Ousterhout's project Magic allows for quick verification of transistor designs so that engineers can easily test multiple designs synthesized by basic concepts \cite{Ousterhout1984}. Leland Wilkinson's ``grammar of graphics'' provides structure and formulation on the massive universe of possible graphics \cite{Wilkinson2005}. Timothy G. Mattson's work creates a language of design patterns for parallel algorithms \cite{Mattson2004}. Mike Franklin's Ph.D. thesis explores the possible client-server architecture designs using caching based replication as the main design primitive \cite{Franklin1993}. 
Joe Hellerstein's work on Generalized Search Trees makes it easy to design and test new data structures by providing templates which expose only a few options where designs need to differ \cite{Hellerstein1995,Aoki1998,Aoki1999,Kornacker1997,Kornacker1999,Kornacker1998,Kornacker2003}. 
S. Bing Yao's \cite{Yao1977} and 
Stefan Manegold's \cite{Manegold2002} work on generalized hardware conscious cost models showed that it is possible to synthesize the costs of complex operations from basic access patterns. 
Work on data representation synthesis in programming languages enables synthesis of representations out of small sets of (3-5) existing data structures \cite{Schonberg1979,Schonberg1981,Cohen1993,Smaragdakis1997,Shacham2009,Hawkins2011,Hawkins2012,Loncaric2016,Steindorfer2016}.

\begin{figure*}[ht]
    \hspace{-.4em}
    \includegraphics[scale=1.06, trim={2.3cm 7.6cm 0.25cm 6.9cm}, clip]{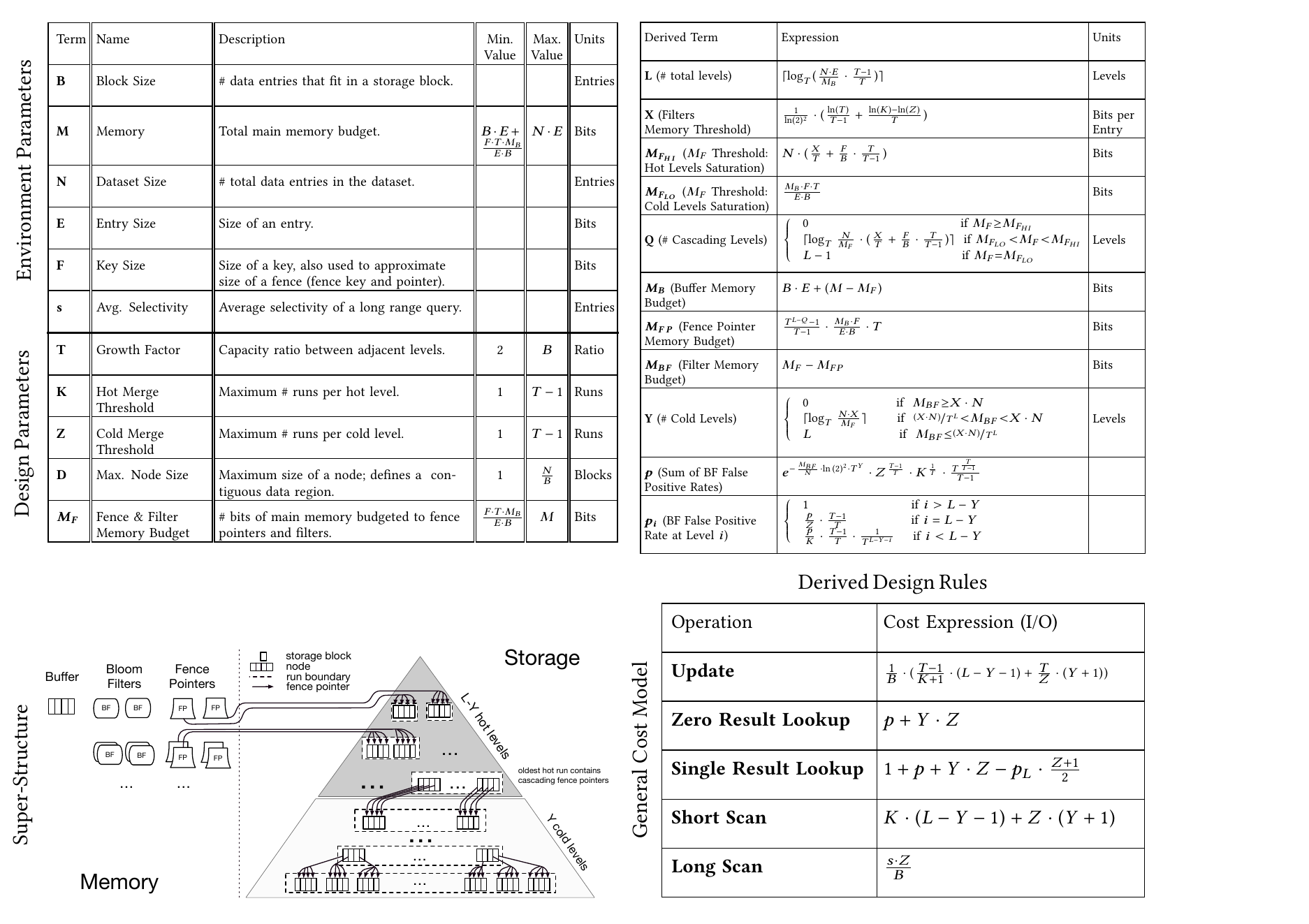}
    \vspace{-2.75em}
        \caption{An example of a design continuum: connecting complex designs with few continuous parameters.}
            \vspace{-0.75em}
            \label{f:continuum}
\end{figure*}

\section{Design Continuums}
\label{sec:design}

We now describe how to construct a design continuum.

\subsection{From B$^{+}$tree to LSM-tree} \label{subsection:originalDC}
We first give an example of a design continuum that connects diverse designs including Tiered LSM-tree \cite{Jagadish1997, Dayan2017, Lakshman2010}, Lazy Leveled LSM-tree \cite{Dayan2018}, Leveled LSM-tree \cite{ONeil1996, Dayan2017, FacebookRocksDB, GoogleLevelDB}, COLA \cite{Bender2007, Jermaine2007}, FD-tree \cite{Li2010}, B$^{\epsilon}$tree \cite{Brodal2003, Arge2003, Bender2007, Jannen2015, Jermaine2007, Papagiannis2016}, and B$^{+}$tree \cite{Bayer1970}. The design continuum can be thought of as a \textbf{super-structure} that encapsulates all those designs. This super-structure consists of $L$ levels where the larger $Y$ levels are \textit{cold} and the smaller $L-Y$ levels are \textit{hot}. Hot levels use in-memory fence pointers and Bloom filters to facilitate lookups, whereas cold levels apply fractional cascading to connect runs in storage. Each level contains one or more runs, and each run is divided into one or more contiguous nodes. There is a write buffer in memory to ingest application updates and flush to Level 1 when it fills up. This overall abstraction allows instantiating any of the data structure designs in the continuum.
Figure \ref{f:continuum} formalizes the continuum and the super-structure is shown at the bottom left. 
For reference, we provide several examples of super-structure instantiations in Appendix \ref{appendix:instances} and Figure \ref{fig:instances}.

\textbf{Environmental Parameters.} The upper right table in Figure \ref{f:continuum} opens with a number of environmental parameters such as dataset size, main memory budget, etc. which are inherent to the application and context for which we want to design a key-value store. 

\textbf{Design Parameters.} The upper right table in Figure \ref{f:continuum} further consists of five continuous design knobs which have been chosen as the smallest set of movable design abstractions that we could find to allow differentiating among the target designs in the continuum. The first knob is the \textit{growth factor} $T$ between the capacities of adjacent levels of the structure (e.g., ``fanout'' for B$^{+}$tree or ``size ratio'' for LSM-tree). This knob allows us to control the super-structure's depth. The second knob is the \textit{hot merge threshold} $K$, which is defined as the maximum number of independent sorted partitions (i.e., runs) at each of Levels 1 to $L-Y-1$ (i.e., all hot levels but the largest) before we trigger merging. The lower we set $K$, the more greedy merge operations become to enforce fewer sorted runs at each of these hot levels. Similarly, the third knob is the \textit{cold merge threshold} $Z$ and is defined as the maximum number of runs at each of Levels $L-Y$ to $L$ (i.e., the largest hot level and all cold levels) before we trigger merging. The \textit{node size} $D$ is the maximal size of a contiguous data region (e.g., a ``node'' in a B$^{+}$tree or ``SSTable'' in an LSM-tree) within a run. Finally, the \textit{fence and filters memory budget} $M_F$ controls the amount of the overall memory that is allocated for in-memory fence pointers and Bloom filters. 

Setting the domain of each parameter is a critical part of crafting a design continuum so we can reach the target designs and correct hybrid designs. Figure \ref{f:continuum} describes how each design parameter in the continuum may be varied. 
For example, we set the maximum value for the size ratio $T$ to be the block size $B$. This ensures that when fractional cascading is used at the cold levels, a parent block has enough space to store pointers to all of its children. As another example, we observe that a level can have at most $T-1$ runs before it runs out of capacity and so based on this observation we set the maximum values of $K$ and $Z$ to be $T-1$.  

\textbf{Design Rules: Forming the Super-structure.} The continuum contains a set of design rules, shown on the upper right part of Figure \ref{f:continuum}. These rules enable instantiating specific designs by deterministically deriving key design aspects. Below we describe the design rules in detail. 

\begin{figure*}[t]
    \includegraphics[scale=1.3, width=\textwidth, trim={3.3cm 10cm 3.2cm 9cm}, clip]{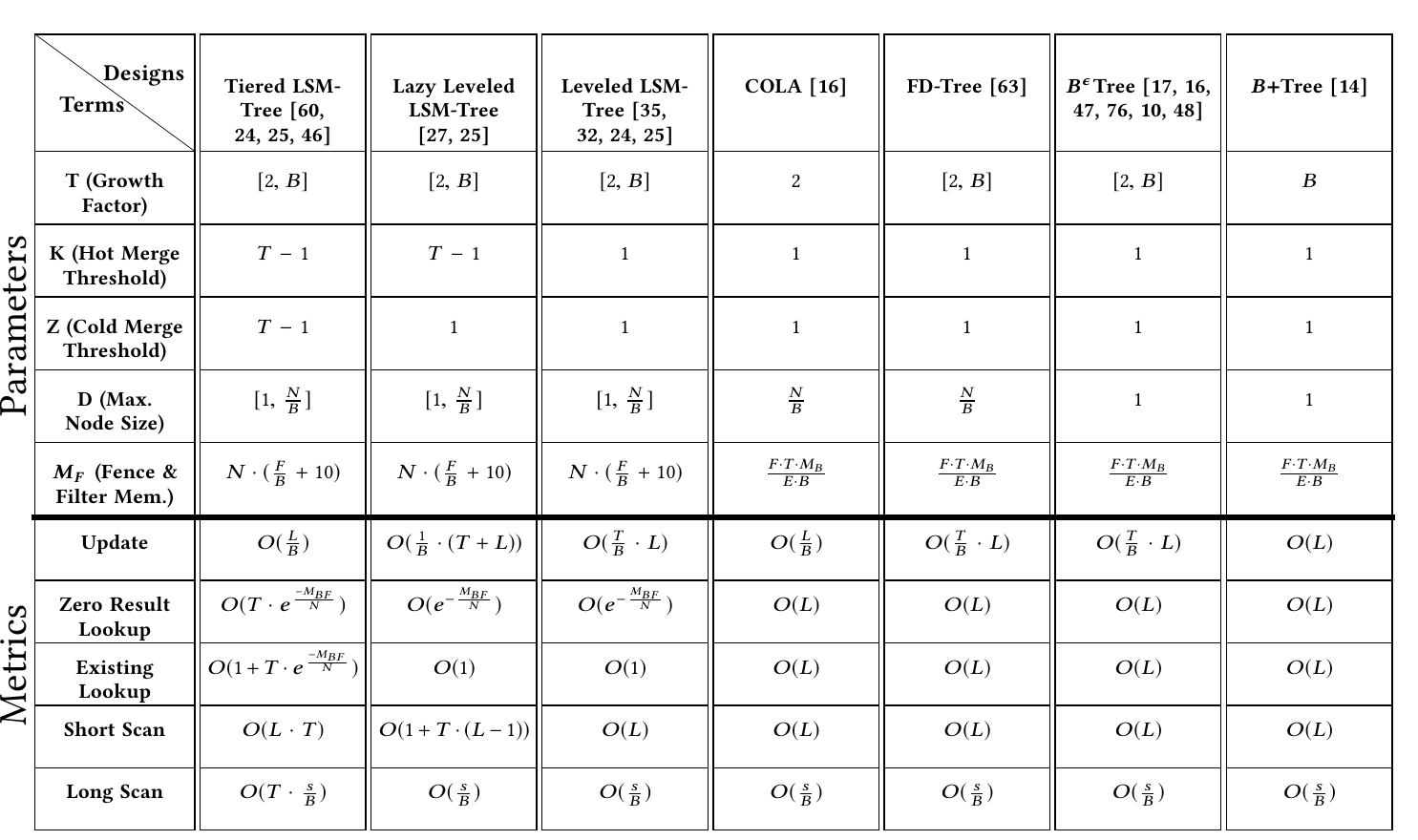}
    \vspace{-1.25em}
    \caption{Instances of the design continuum and examples of their derived cost metrics.}
        \vspace{-1.5em}
    \label{f:instances}
\end{figure*}

\textit{Exponentially Increasing Level Capacities.}
The levels' capacities grow exponentially by a factor of $T$ starting with the write buffer's capacity. As a result, the overall number of levels $L$ grows logarithmically with the data size. 


\textit{Fence Pointers \& Bloom Filters.}
Our design allocates memory for fence pointers and Bloom filters from smaller to larger levels based on the memory budget assigned by the knob $M_F$. Specifically, we first compute $Q$ as the number of levels for which there is not enough memory for having both Bloom filters and fence pointers. 
We then assign the memory budget $M_F$ for fence pointers to as many levels as there is enough memory for. This is shown by the Equation for the fence pointers budget $M_{FP}$ in Figure \ref{f:continuum}. The remaining portion of $M_F$ after fence pointers is assigned to a Bloom filters memory budget $M_{BF}$. This can also be done in the reverse way when one designs a structure, i.e., we can define the desired write buffer budget first and then give the remaining from the total memory budget to filters and fence pointers. 

\textit{Optimal Bloom Filter Allocation Across Levels.}
The continuum assigns exponentially decreasing false positive rates (FPRs) to Bloom filters at smaller levels, as this approach was shown to minimize the sum of their false positive rates and thereby minimize point read cost \cite{Dayan2017}. In Figure \ref{f:continuum}, we express the FPR assigned to Level $i$ as $p_i$ and give corresponding equations for how to set $p_i$ optimally with respect to the different design knobs.



\textit{Hot vs. Cold Levels.}
Figure \ref{f:continuum} further shows how to compute the number of cold levels $Y$ for which there is not sufficient memory for Bloom filters. The derivation for $Y$ is in terms of a known threshold $X$ for when to drop a filter for a level and instead use that memory for filters at smaller levels to improve performance \cite{Dayan2018}. Note that $Y$ is either equal to $Q$ or larger than it by at most one.
We derive $M_{F_{HI}}$ as the amount of memory above which all levels are hot (i.e., $Y=0$). We also set a minimum memory requirement $M_{F_{LO}}$ on $M_F$ to ensure that there is always enough memory for fence pointers to point to Level 1.

\textit{Fractional Cascading for Cold Levels.}
We use fractional cascading to connect data at cold levels to the structure, to address the issue of not having enough memory to point to them using in-memory fence pointers. For every block within a run at a cold level, we embed a ``cascading'' fence pointer within the next younger run along with the smallest key in the target block. This allows us to traverse cold levels with one I/O for each run by following the corresponding cascading fence pointers to reach the target key range. 

\textit{Active vs. Static Runs.} Each level consists of one \textit{active run} and a number of \textit{static runs}. Incoming data into a level gets merged into the active run. When the active run reaches a fraction of {{\small $\nicefrac{T}{K}$}} of the levels' capacity for Levels 1 to {\small $L-Y-1$} or {{\small$\nicefrac{T}{Z}$}} for Levels {\small $L-Y$} to {\small $L$}, it becomes a static run and a new empty active run is initialized. 

\textit{Granular Rolling Merging.} When a level reaches capacity, a merge operation needs to take place to free up space. 
We perform a merge by first picking an \textit{eviction key}
\footnote{In practice, an eviction key implies a range of eviction keys with a start key and end key, because across runs the notion of temporal sequentiality for entries with the same key must be maintained. The strategy for picking the eviction key may be as simple as round robin, though more sophisticated strategies to minimize key overlap with the active run in the next level are possible so as to minimize merge overheads \cite{Thonangi2017}. }. 
Since each run is sorted across its constituent nodes, there is at most one node in each of the static runs at the level that intersects with the eviction key. 
We add these nodes into an \textit{eviction} set and merge them into the active run in the next larger level. 
Hence, the \textit{merge granularity} is controlled by the maximum node size $D$, and merge operations \textit{roll} across static runs and eventually empty them out. 

\textit{Fractional Cascading Maintenance.} As merge operations take place at cold levels, cascading fence pointers must be maintained to keep runs connected. As an active run gradually fills up, we must embed cascading fence pointers from within the active run at the next smaller level. We must also create cascading fence pointers from a new active run into the next older static run at each level. 
To manage this, whenever we create a new run, we also create a \textit{block index} in storage to correspond to the fences for this new run. Whenever we need to embed pointers into a Run $i$ from some new Run $j$ as Run $j$ is being created, we include the block index for Run $i$ in the sort-merge operation used to create Run $j$ to embed the cascading fence pointers within the constituent nodes of that run.

\textbf{Unified Cost Model.} A design continuum includes a cost model with a closed-form equation for each one of the core performance metrics. The bottom right part of Figure \ref{f:continuum} depicts these models for our example continuum. These cost models measure the worst-case number of I/Os issued for each of the operation types, the reason being that I/O is typically the performance bottleneck for key-value stores that store a larger amount of data than can fit in memory.\footnote{Future work can also try to generate in-memory design continuums where we believe learned cost models that help synthesize the cost of arbitrary data structure designs can be a good start  \cite{Idreos2018}.} For example, the cost for point reads is derived by adding the expected number of I/Os due to false positives across the hot levels (given by the Equation for $p$, the sum of the FPRs \cite{Dayan2018}) to the number of runs at the cold levels, since with fractional cascading we perform 1 I/O for each run. As another example, the cost for writes is derived by observing that an application update gets copied on average $O(\nicefrac{T}{K})$ times at each of the hot levels (except the largest) and $O(\nicefrac{T}{Z})$ times at the largest hot level and at each of the cold levels. 
We amortize by adding these costs and dividing by the block size $B$ as a single write I/O copies $B$ entries from the original runs to the resulting run. 

While our models in this work are expressed in terms of asymptotic notations, we have shown in earlier work that such models can be captured more precisely to reliably predict worst-case performance \cite{Dayan2017, Dayan2018}. A central advantage of having a set of closed-form set of models is that they allow us to see how the different knobs interplay to impact performance, and they reveal the trade-offs that the different knobs control.  

Overall, the choice of the design parameters and the derivation rules represent the infusion of expert design knowledge in order to create a navigable design continuum. Specifically, \textbf{fewer design parameters} (for the same target designs) lead to a cleaner abstraction which in turn makes it easier to come up with algorithms that automatically find the optimal design (to be discussed later on). We minimize the number of design parameters in two ways: 1) by adding deterministic design rules which encapsulate expert knowledge about what is a good design, and 2) by collapsing more than one interconnected design decisions to a single design parameter. 
For example, we used a single parameter for the memory budget of Bloom filters and fence pointers as we assume their co-existence since fence pointers accelerate both point lookups and range queries, while Bloom filters accelerate point lookups exclusively. We specify a design strategy that progressively adds more point query benefit as the memory for fences and filters increases by prioritizing fence budget before filter budget at each level, accordingly jointly budgeting fences and filters together as one design knob. 

\begin{figure*}[t]
    \includegraphics[width=\textwidth]{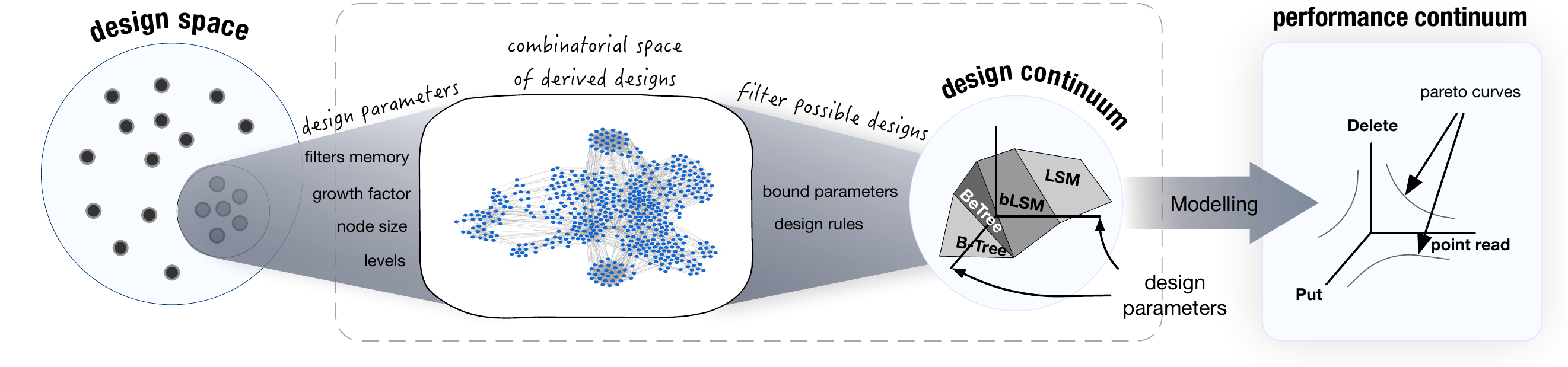}
        \vspace{-2.5em}
    \caption{Constructing a design continuum: from design parameters to a performance hyperplane.}
            \vspace{-1em}
    \label{f:construction}
\end{figure*}




\textbf{Design Instances.} Figure \ref{f:instances} depicts several known instances of data structure designs as they are derived from the continuum. 
In particular, the top part of Figure \ref{f:instances} shows the values for the design knobs that derive each specific design, and the bottom part shows how their costs can indeed be derived from the generalized cost model of the continuum. 	
Additional example diagrams of super-structure instantiations are provided in Appendix \ref{appendix:instances} Figure \label{fig:instances}

For example, a B$^{+}$tree is instantiated by (1) setting the maximum node size $D$ to be one block\footnote{Node size can be set to whatever we want the B$^{+}$tree node size to be - we use $D=1$ block here as an example only.}, (2) setting $K$ and $Z$ to 1 so that all nodes within a level are globally sorted, (3) setting $M_F$ to the minimum amount of memory so that Levels 1 to L get traversed using fractional cascading without the utilization of Bloom filters or in-memory fence pointers, and (4) setting the growth factor to be equal to the block size. By plugging the values of these knobs into the cost expressions, the well-known write and read costs for a B$^{+}$tree of $O(L)$ I/Os immediately follow. 

As a second example, a leveled LSM-tree design is instantiated by (1) setting $K$ and $Z$ to 1 so that there is at most one run at each level, and (2) assigning enough memory to the knob $M_F$ to enable fence pointers and Bloom filters (with on average 10 bits per entry in the table) for all levels. We leave the knobs $D$ and $T$ as variables in this case as they are indeed used by modern leveled LSM-tree designs to strike different trade-offs. By plugging in the values for the design knobs into the cost models, we immediately obtain the well-known costs for a leveled LSM-tree. For example, write cost simplifies to $O(\frac{T \cdot L}{B})$ as every entry gets copied across $O(L)$ levels and on average $O(T)$ times within each level.

\textbf{Construction Summary.} Figure \ref{f:construction} summarizes the process of constructing a design continuum. We start by selecting a set of data structures. Then we select the minimum set of design knobs that can instantiate these designs and we impose design rules and domain restrictions to restrict the population of the continuum to only the best designs with respect to our target cost criteria. Finally, we derive the generalized cost models. 

\textbf{Definition of Continuum.} We can now revisit the exact definition of the continuum. A design continuum connects previously distinct and seemingly fundamentally different data structure designs. The construction process does not necessarily result in continuous knobs in the mathematical sense (most of the design knobs have integer values). However, from a design point of view a continuum opens the subspace in between previously unconnected designs; it allows us to connect those discrete designs in fine grained steps, and this is exactly what we refer to as the ``design continuum''. The reason that this is critical is that it allows us to 1) ``see'' designs that we did not know before, derived as combinations of those fine-grained design options, and 2) build techniques that smoothly transition across discrete designs by using those intermediate states.  
 
\subsection{Interactive Design}
The generalized cost models enable us to navigate the continuum, i.e., interactively design a data structure for a key-value store with the optimal configuration for a particular application as well as to react to changes in the environment, or workload. We formalize the navigation process by introducing Equation \ref{eq:avg_cost1} to model the average operation cost $\theta$ through the costs of zero-result point lookups $R$, non-zero-result point lookups $V$, short range lookups $Q$, long range lookups $C$, and updates $W$ (the coefficients depict the proportions of each in the workload).  
\vspace{-.4em}
\begin{equation} \label{eq:avg_cost1}
\small
\theta = ( r \cdot R + v \cdot V + q \cdot Q + c \cdot C + w \cdot W  )
\end{equation}

\vspace{-.4em}
To design a data structure using Equation \ref{eq:avg_cost1}, we first identify the bottleneck as the highest additive term as well as which knobs in Figure \ref{f:continuum} can be tweaked to alleviate it. We then tweak the knob in one direction until we reach its boundary or until $\theta$ reaches the minimum with respect to that parameter. We then repeat this process with other parameters as well as with other bottlenecks that can emerge during the process. This allows us to converge to the optimal configuration without backtracking, which allows us to  adjust to a variety of application scenarios reliably. For example, consider an application with a workload consisting of point lookups and updates and an initial configuration of a lazy-leveled LSM-tree with {\small$T=10$}, {\small$K=T-1$}, {\small$Z=1$}, {\small$D=64$}, {\small$M_B$} set  to 2 MB, and {\small$M_f$} set to {\small$N\cdot (F/B+10)$}, meaning we have memory for all the fence pointers and in addition 10 bits per entry for Bloom filters. We can now use the cost models to react to different scenarios. 

\emph{Scenario 1: Updates Increasing.}  Suppose that the proportion of updates increases, as is the case for many applications \cite{Sears2012}. To handle this, we first increase $Z$ until we reach the minimum value for $\theta$ or until we reach the maximum value of $Z$. If we reach the maximum value of $Z$, the next promising parameter to tweak is the size ratio $T$, which we can increase in order to decrease the number of levels across which entries get merged. Again, we increase $T$ until we hit its maximum value or reach a minimum value for $\theta$. 

\emph{Scenario 2: Range Lookups.} Suppose that the application changes such that short-range lookups appear in the workload. To optimize for them, we first decrease $K$ to restrict the number of runs that lookups need to access across Levels 1 to {\small$L-1$}. If we reach the minimum value of $K$ and short-range lookups remain the bottleneck, we can now increase $T$ to decrease the overall number of levels  thereby decreasing the number of runs further. 

\emph{Scenario 3: Data Size Growing.} Suppose that the size of the data is growing, yet most of the lookups are targeting the youngest $N_{youngest}$ entries, and we do not have the resources to continue scaling main memory in proportion to the overall data size $N$. In such a case, we can fix $M_f$ to  $N_{youngest} \cdot (F/B+10)$ to ensure memory is invested to provide fast lookups for the hot working set while minimizing memory overhead of less frequently requested data by maintaining cold levels with fractional cascading. 

Effectively the above process shows how to quickly and reliably go from a high-level workload requirement to a low-level data structure design configuration at interactive times using the performance continuum. 

\textbf{Auto-Design.} It is possible to take the navigation process one step further to create algorithms that iterate over the continuum and independently find the best configuration. The goal is to find the best values for $T$, $K$, $Z$, $D$, and the best possible division of a memory budget between $M_F$ and $M_B$. While iterating over every single configuration would be intractable as it would require traversing every permutation of the parameters, we can leverage the manner in which we constructed the continuum to significantly prune the search space. For example, in Monkey \cite{Dayan2017}, when studying a design continuum that contained only a limited set of LSM-tree variants we observed that two of the knobs have a logarithmic impact on $\theta$, particularly the size ratio $T$ and the memory allocation between $M_b$ and $M_f$. For such knobs, it is only meaningful to examine a logarithmic number of values that are exponentially increasing, and so their multiplicative contribution to the overall search time is logarithmic in their domain. 
While the continuum we showed here is richer, by adding B-tree variants, this does not add significant complexity in terms of auto-design. The decision to use cascading fence pointers or in-memory fence pointers completely hinges on the allocation of memory between $M_F$ and $M_B$, while the node size $D$ adds one multiplicative logarithmic term in the size of its domain.


\subsection{Success Criteria} 

We now outline the ideal success criteria that should guide the construction of elegant and practically useful design continuums in a principled approach. 

\begin{figure*}[t]
    \includegraphics[scale=1.02, width=\textwidth, trim={1.75cm 12.5cm 3.25cm 11.5cm}, clip]{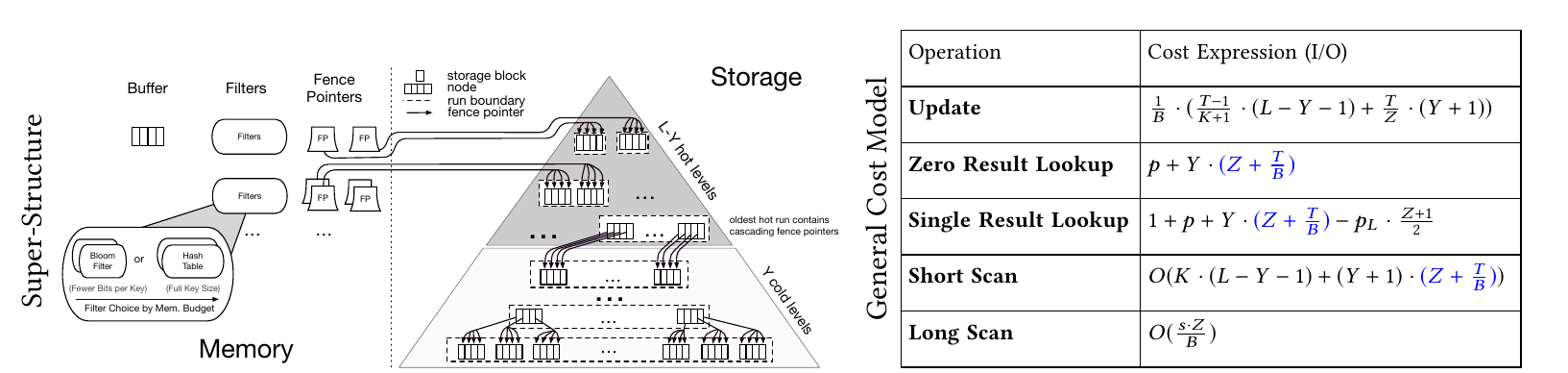}
    \vspace{-1.75em}
     \caption{Extending the design continuum to support Log Structured Hash table designs.}
    \vspace{-1.25em}
    \label{f:modellsh}
\end{figure*}

\textbf{Functionally Intact.} All possible designs that can be assumed by a continuum should be able to correctly support all operation types (e.g., writes, point reads, etc.). In other words, a design continuum should only affect the performance properties of the different operations rather than the results that they return. 

\textbf{Pareto-Optimal.} All designs that can be expressed should be Pareto-optimal with respect to the cost metrics and workloads desired. This means that there should be no two designs such that one of them is better than the other on one or more of the performance metrics while being equal on all the others. The goal of only supporting Pareto-optimal designs is to shrink the size of the design space to the minimum essential set of knobs that allow to control and navigate across only the best possible known trade-offs, while eliminating inferior designs from the space. 

\textbf{Bijective.} A design continuum should be a bijective (one-to-one) mapping from the domain of design knobs to the co-domain of performance and memory trade-offs. As with Pareto-Optimality, the goal with bijectivity is to shrink a design continuum to the minimal set of design knobs such that no two designs that are equivalent in terms of performance can be expressed as different knob configurations. If there are multiple designs that map onto the same trade-off, it is a sign that the model is either too large and can be collapsed onto fewer knobs, or that there are core metrics that we did not yet formalize, and that we should. 

\textbf{Diverse.} A design continuum should enable a diverse set of performance properties. For Pareto-Optimal and bijective continuums, trade-off diversity can be measured and compared across different continuums as the product of the domains of all the design knobs, as each unique configuration leads to a different unique and Pareto-optimal trade-off.

\textbf{Navigable.} The time complexity required for navigating the continuum to converge onto the optimal (or even near-optimal) design should be tractable. With the Monkey continuum, for example, we showed that it takes {\small $O(log_T(N))$} iterations to find the optimal design \cite{Dayan2017}, and for Dostoevsky, which includes more knobs and richer trade-offs, we showed that it takes {\small $O(log_T(N)^3)$} iterations \cite{Dayan2018}. Measuring navigability complexity in this way allows system designers from the onset to strike a balance between the diversity vs. the navigability of a continuum. 

\textbf{Layered.} By construction, a design continuum has to strike a trade-off between diversity and navigability. The more diverse a continuum becomes through the introduction of new knobs to assume new designs and trade-offs, the longer it takes to navigate it to optimize for different workloads. With that in mind, however, we observe that design continuums may be constructed in layers, each of which builds on top of the others. Through layered design, different applications may use the same continuum but choose the most appropriate layer to navigate and optimize performance across. For example, the design continuum in Dostoevsky \cite{Dayan2017} is layered on top of Monkey \cite{Dayan2018} by adding two new knobs, $K$ and $Z$, to enable intermediate designs between tiering, leveling and lazy leveling. While Dostoevsky requires {\small $O(log_T(N)^3)$} iterations to navigate the possible designs, an alternative is to leverage layering to restrict the knobs $K$ and $Z$ to both always be either 1 or $T-1$ (i.e., to enable only leveling and tiering) in order to project the Monkey continuum and thereby reduce navigation time to {\small $O(log_T(N))$}. In this way, layered design enables \textit{continuum expansion with no regret}: we can continue to include new designs in a continuum to enable new structures and trade-offs, all without imposing an ever-increasing navigation penalty on applications that need only some of the possible designs.

\subsection{Expanding a Continuum: \\ A Case-Study with LSH-table} \label{subsection:extendedDC}
We now demonstrate how to expand the continuum with a goal of adding a particular design to include certain performance trade-offs. The goal is to highlight the design continuum construction process and principles. 

Our existing continuum does not support the LSH-table data structure used in many key-value stores such as BitCask \cite{Sheehy2010}, FASTER \cite{Chandramouli2018}, and others \cite{Ahn2016, Lim2014, Rumble2014, SpotifySparkey, Wu2018}. LSH-table achieves a high write throughout by logging entries in storage, and it achieves fast point reads by using a hash table in memory to map every key to the corresponding entry in the log.  In particular, LSH-table supports writes in $O(1/B)$ I/O, point reads in $O(1)$ I/O, range reads in {\small $O(N)$} I/O, and it requires {\small$O(F \cdot N)$} bits of main memory to store all keys in the hash table. As a result, it is suitable for write-heavy application with ample memory, and no range reads.

We outline the process of expanding our continuum in three steps: \textit{bridging, patching, and costing}.

\textbf{Bridging.} Bridging entails identifying the least number of new movable design abstractions to introduce to a continuum  to assume a new design. This process involves three options: 1) introducing new design rules, 2) expanding the domains of existing knobs, and 3) adding new design knobs.  

Bridging increases the diversity of a design continuum, though it risks compromising the other success metrics. Designers of continuums should experiment with the three steps above in this particular order to minimize the chance of that happening. With respect to LSH-table, we need two new abstractions: one to allow assuming a log in storage, and one to allow assuming a hash table in memory. 

To assume a log in storage, our insight is that with a tiered LSM-tree design, setting the size ratio to increase with respect to the number of runs at Level 1 (i.e., {\small $T= \nicefrac{(N \cdot E)}{M_B}$}) causes Level~1 to never run out of capacity. This effectively creates a log in storage as merge operations never take place. Our current design continuum, however, restricts the size ratio to be at most $B$. To support a log, we expand the domain of the size ratio with a new maximum value of {\small $\nicefrac{(N \cdot E)}{M_B}$}. 

\begin{figure}[t]
    \includegraphics[scale=1.05, width=\textwidth, trim={1.75cm 9.5cm 1.75cm 8.5cm}, clip]{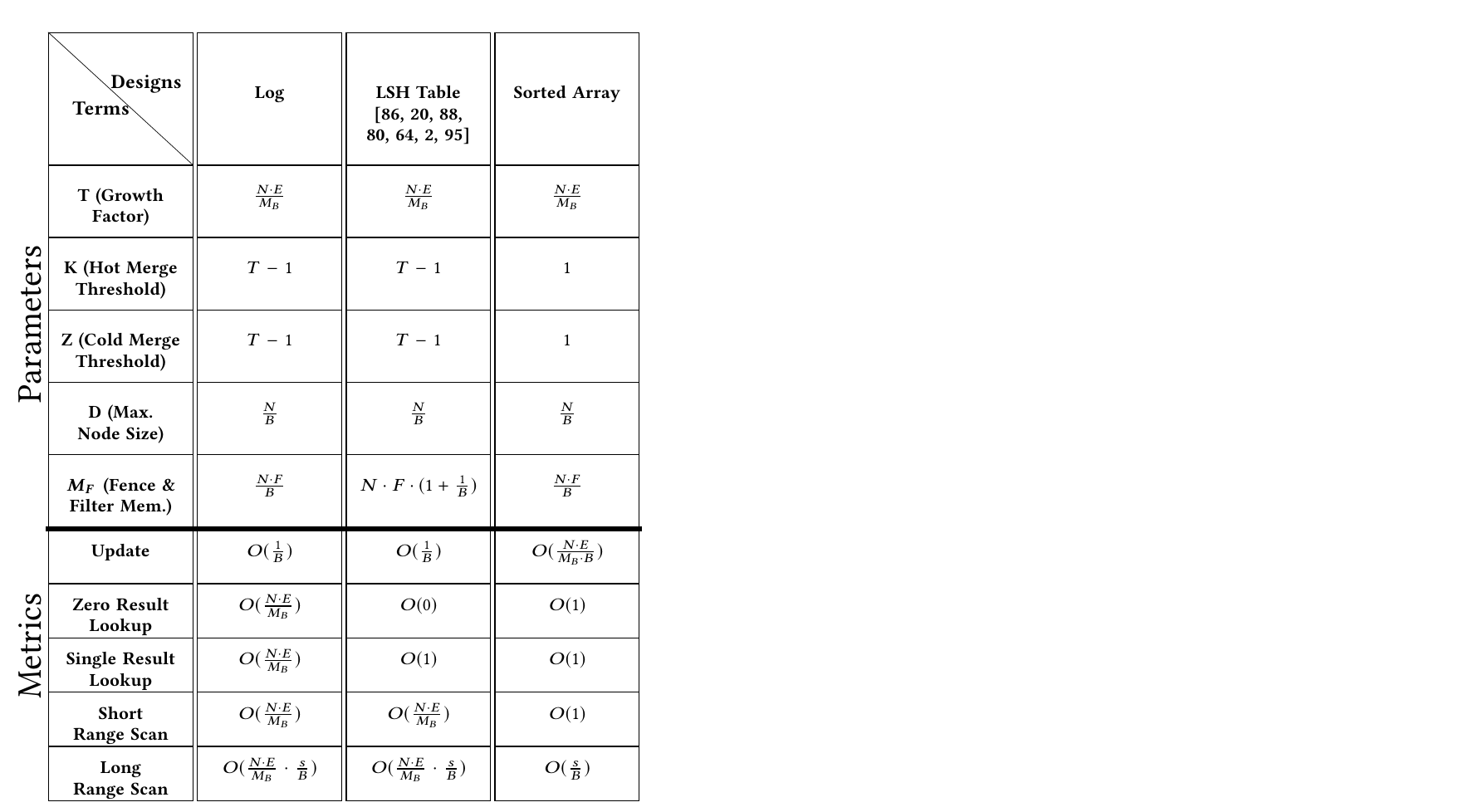}
    \vspace{-2em}
    \caption{Instances of the extended design continuum and examples of their derived cost metrics.}
    \label{f:instanceslsh}
\end{figure}

To assume a hash table in memory, recall that our continuum assigns more bits per entry for Bloom filters at smaller levels. Our insight is that when the number of bits per entry assigned to given level exceeds the average key size $F$, it is always beneficial to replace the Bloom filters at that level with an in-memory hash table that contains all keys at the level. The reason is that a hash table takes as much  memory as the Bloom filters would, yet it is more precise as it does not allow false positives at all. We therefore introduce a new design rule whereby levels with enough memory to store all keys use a hash table while levels with insufficient memory use Bloom filters\footnote{Future work toward an even more navigable continuum can attempt to generalize a Bloom filter and a hash table into one unified model with continuous knobs that allows to gradually morph between these structures based on the amount of main memory available.}. With these two new additions to the continuum, we can now set the size ratio to $\nicefrac{(N \cdot E)}{M_B}$ and $K$ and $Z$ to {\small $T-1$} while procuring at least {\small $F \cdot N$} bits of memory to our system to assume LSH-table\footnote{More precisely, { $F \cdot N \cdot (1 + \frac{1}{B})$ bits of memory are needed to support both the hash table and fence pointers.}}. Figure \ref{f:modellsh} shows the new super-structure of the continuum while Figure \ref{f:instanceslsh} shows how LSH-table can be derived. 

An important point is that we managed to bridge LSH-table with our continuum without introducing new design knobs. As a rule of thumb, introducing new knobs for bridging should be a last resort as the additional degrees of freedom increase the time complexity of navigation. Our case-study here, however, demonstrates that even data structures that seem very different at the onset can be bridged by finding the right small set of movable abstractions. 


\textbf{Patching.} Since the bridging process introduces many new intermediate designs, we follow it with a patching process to ensure that all of the new designs are functionally intact (i.e., that they can correctly support all needed types of queries). Patching involves either introducing new design rules to fix broken designs or adding constraints on the domains of some of the knobs to eliminate broken designs from the continuum. To ensure that the expanded continuum is layered (i.e., that it contains all designs from the continuum that we started out with), any new design rules or constraints introduced by patching should only affect new parts of the continuum. Let us illustrate an example of patching with the expanded continuum. 

The problem that we identify arises when fractional cascading is used between two cold Levels $i$ and $i+1$ while the size ratio $T$ is set to be greater than $B$. In this case, there is not enough space inside each block at Level $i$ to store all pointers to its children blocks (i.e., ones with an overlapping key range) at Level $i+1$. The reason is that a block contains $B$ slots for pointers, and so a block at Level $i$ has a greater number of children $T$ than the number of pointer slots available. Worse, if the node size $D$ is set to be small (in particular, when {\small $D < \nicefrac{T}{B}$}), some of the blocks at Level $i+1$ will neither be pointed to from Level $i$ nor exist within a node whereon at least one other block is pointed to from Level $i$. As a result, such nodes at Level  $i+1$ would \textit{leak} out of the data structure, and so the data on these blocks would be lost. To prevent leakage, we introduce a design rule that when {\small $D < \nicefrac{T}{B}$} and {\small $B < T$}, the setting at which leakage can occur, we add sibling pointers to reconnect nodes that have leaked. We introduce a rule that the parent block's pointers are spatially evenly distributed across its children (every {\small $(\nicefrac{T}{(B \cdot D)})^{\text{th}}$} node at Level $i+1$ is pointed to from a block at level $i$) to ensure that all sibling chains of nodes within Level $i+1$ have an equal length. As these new rules only apply to new parts of our continuum (i.e., when {\small $T > B$)}, they do not violate layering.

\textbf{Costing.} The final step is to generalize the continuum's cost model to account for all new designs. This requires either extending the cost equations and/or proving that the existing equations still hold for the new designs. Let us illustrate two examples. First, we extend the cost model with respect to the patch introduced above. In particular, the lookup costs need to account for having to traverse a chain of sibling nodes at each of the cold levels when $T > B$. As the length of each chain is $\nicefrac{T}{B}$ blocks, we extend the cost equations for point lookups and short-range lookups with additional $\nicefrac{T}{B}$ I/Os per each of the $Y$ cold levels. The extended cost equations are shown in Figure \ref{f:modellsh}.  

In the derivation below, we start with general cost expression for point lookups in Figure \ref{f:modellsh} and show how the expected point lookup cost for LSH-table is indeed derived correctly. In Step \ref{eq:step1}, we plug in $\nicefrac{N}{B}$ for $T$ and $Z$ to assume a log in storage. In Step \ref{eq:step2}, we set the number of cold levels to zero as Level 1 in our continuum by construction is always hot and in this case, there is only one level (i.e., $L=1$), and thus $Y$ must be zero. In Step \ref{eq:step3}, we plug in the key size $F$ for the number of bits per entry for the Bloom filters, since with LSH-table there is enough space to store all keys in memory. In Step \ref{eq:step4}, we reason that the key size $F$ must comprise on average at least $log(N)$ bits to represent all unique keys. In Step \ref{eq:step5}, we simplify and omit small constants to arrive at a cost of $O(1)$ I/O per point lookup. 

{
	\scriptsize
	\begin{align}
	&\in   O( 1 + Z \cdot e^{ - (\nicefrac{M_{BF}}{N}) \cdot T^Y } + Y \cdot (Z + \nicefrac{T}{B}) )    &&  \nonumber \\ 
	&\in O( 1 + \nicefrac{N}{B} \cdot e^{ - (\nicefrac{M_{BF}}{N}) \cdot (\nicefrac{N}{B})^Y } + Y \cdot (\nicefrac{N}{B} + \nicefrac{N}{B^2}) ) &&    \label{eq:step1} \\
	&\in O( 1 + \nicefrac{N}{B} \cdot e^{ - (\nicefrac{M_{BF}}{N}) }  ) &&   \label{eq:step2} \\
	&\in O( 1 + \nicefrac{N}{B} \cdot e^{ - F }  ) &&   \label{eq:step3} \\
	&\in O( 1 + \nicefrac{N}{B} \cdot e^{ - \log_2(N) }  ) &&  \label{eq:step4} \\
	&\in O( 1 )   &&  \label{eq:step5} 
	\end{align}
}

\subsection{Elegance Vs. Performance: \\ To Expand or Not to Expand?}

As new data structures continue to get invented and optimized, the question arises of when it is desirable to expand a design continuum to include a new design. We show through an example that the answer is not always clear cut.

In an effort to make B-trees more write-optimized for flash devices, several recent B-tree designs buffer updates in memory and later flush them to a log in storage in their arrival order. They further use an in-memory indirection table to map each logical B-tree node to the locations in the log that contain entries belonging to that given node. This design can improve on update cost relative to a regular B-tree by flushing multiple updates that target potentially different nodes with a single sequential write. The trade-off is that during reads, multiple I/Os need to be issued to the log for every logical B-tree node that gets traversed in order to fetch its contents. To bound the number of I/Os to the log, a compaction process takes place once a logical node spans over $C$ blocks in the log, where $C$ is a tunable parameter. Overall, this design leads to a point and range read cost of {\small$O(C \cdot \log_B(N))$} I/Os. On the other hand, update cost consists of {\small$O(C \cdot \log_B(N))$} read I/Os to find the target leaf node and an additional amortized  {\small $O(\nicefrac{1}{C})$} write I/Os to account for the overheads of compaction. The memory footprint for the mapping table is {\small$O(\nicefrac{(C \cdot N \cdot F)}{B})$} bits. We refer to this design as log-structured B-tree (LSB-tree). Would we benefit from including LSB-tree in our continuum? 

\begin{figure}[t]
	\centering
	\includegraphics[width=0.9\linewidth]{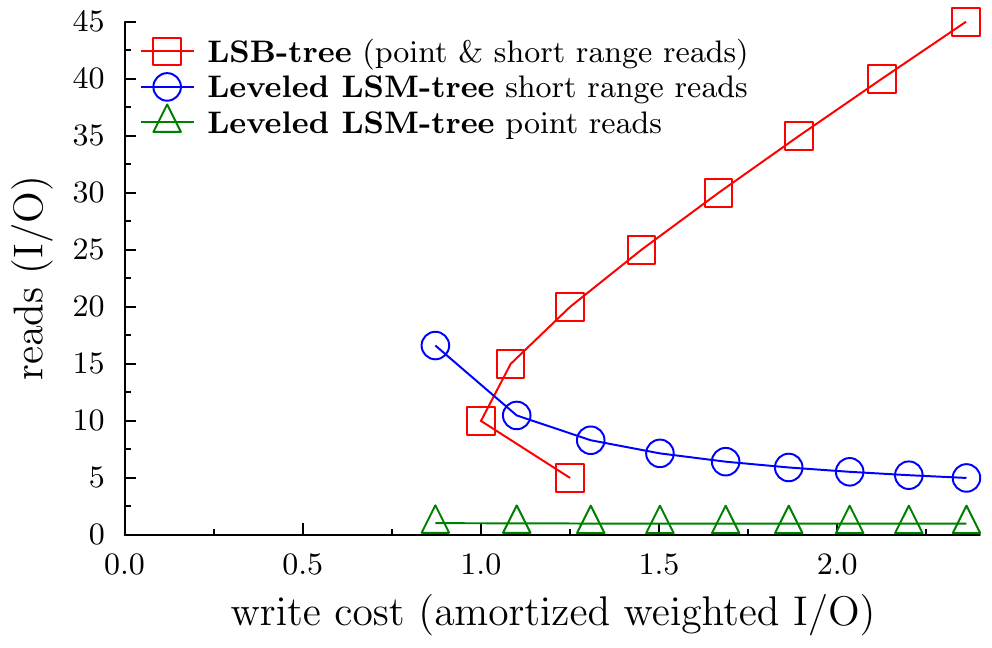}
	\caption{Leveled LSM-tree dominates LSB-tree for most of their respective continuums.}
	\label{fig:bw}
\end{figure}

To approach an answer to this question, we analytically compare LSB-tree against designs within our continuum to gauge the amount by which LSB-tree would allow us to achieve better trade-offs with respect to our continuum's cost metrics. We demonstrate this process in Figure \ref{fig:bw}, which plots point and range read costs against write cost for both LSB-tree and Leveled LSM-tree, a representative part of our continuum. To model write cost for LSB-tree, we computed a weighted cost of {\small$O(C \cdot \log_B(N))$} read I/Os to traverse the tree, {\small$O(\nicefrac{1}{C})$} write I/Os to account for compaction overheads, and we discounted the cost of a read I/O relative to a write I/O by a factor of 20 to account for read/write cost asymmetries on flash devices. We generated the curve for LSB-tree by varying the compaction factor $C$ from 1 to 9, and the curves for the LSM-tree by varying the size ratio $T$ from 2 to 10. To enable an apples-to-apples comparison whereby both LSB-tree and the LSM-tree have the same memory budget, we assigned however much main memory LSB-tree requires for its mapping table to the LSM-tree's fence pointers and Bloom filters. Overall, the figure serves as a first approximation for the trade-offs that LSB-tree would allow us to achieve relative to our continuum. 

Figure \ref{fig:bw} reveals that point read cost for the LSM-tree is much lower than for LSB-tree. The reason is that when the same amount of memory required by LSB-tree's memory budget is used for the LSM-tree's fence pointers and Bloom filters, hardly any false positives take place and so the LSM-tree can answer most point reads with just one I/O. Secondly, we observe that as we increase LSB-tree's compaction factor $C$, write cost initially decreases but then starts degrading rapidly. The reason is that as $C$ grows, more reads I/Os are required by application writes to traverse the tree to identify the target leaf node for the write. 
On the other hand, for range reads there is a point at which LSB-tree dominates the LSM-tree as fewer blocks need to be accessed when $C$ is small.  

\textbf{Elegance and Navigability versus Absolute Performance.} By weighing the advantages of LSB-tree against the complexity of including it (i.e., adding movable abstractions to assume indirection and node compactions), one can decide to leave LSB-tree out of the continuum. This is because its design principles are fundamentally different than what we had included and so substantial changes would be needed that would complicate the continuum's construction and navigability. On the other hand, when we did the expansion for LSH-table, even though, it seemed initially that this was a fundamentally different design, this was not the case: LSH-table is synthesized from the same design principles we already had in the continuum, and so we could achieve the expansion in an elegant way at no extra complexity and with a net benefit of including the new performance trade-offs.

At the other extreme, one may decide to include LSB-tree because the additional performance trade-offs outweigh the complexity for a given set of desired applications. We did this analysis to make the point of elegance and navigability versus absolute performance. However, we considered a limited set of performance metrics, i.e., worst-case I/O performance for writes, point reads and range reads. Most of the work on LSB-tree-like design has been in the context of enabling better concurrency control \cite{Levandoski2013} and leveraging workload skew to reduce compaction frequency and overheads \cite{Wu2007}. Future expansion of the design space and continuums should include such metrics and these considerations described above for the specific example will be different. In this way, the decision of whether to expand or not to expand a continuum is a continual process, for which the outcome may change over time as different cost metrics change in their level of importance given target applications.

\section{Why not Mutually Exclusive \\ Design Components?} 

Many modern key-value stores are composed of mutually exclusive sets of swappable data layout designs to provide diverse performance properties. For example, WiredTiger supports separate B-tree and LSM-tree implementations to optimize more for reads or writes, respectively, while RocksDB files support either a sorted strings layout or a hash table layout to optimize more for range reads or point reads, respectively. A valid question is how does this compare to the design continuum in general? And in practice how does it compare to the vision of self-designing key-value stores? 

Any exposure of data layout design knobs is similar in spirit and goals to the continuum but how it is done exactly is the key. Mutually exclusive design components can be in practice a tremendously useful tool to allow a single system to be tuned for a broader range of applications than we would have been able to do without this feature. However, it is not a general solution and  leads to three fundamental problems.  

\textbf{1) Expensive Transitions.} Predicting the optimal set of design components for a given application before deployment is hard as the workload may not be known precisely. As a result, components may need to be continually reshuffled during runtime. Changing among large components during runtime is disruptive as it often requires rewriting all data. In practice, the overheads associated with swapping components often force practitioners to commit to a suboptimal design from the onset for a given application scenario. 

\textbf{2) Sparse Mapping to Performance Properties.} An even deeper problem is that mutually exclusive design components tend to have polar opposite performance properties (e.g., hash table vs. sorted array). Swapping between two components to optimize for one operation type (e.g. point reads) may degrade a different cost metric (e.g. range reads) by so much that it would offset the gain in the first metric and lead to poorer performance overall. In other words, optimizing by shuffling components carries a risk of overshooting the target performance properties and hitting the point of diminishing returns. A useful way of thinking about this problem is that mutually exclusive design components map sparsely onto the space of possible performance properties. The problem is that, with large components, there are no intermediate designs that allow to navigate performance properties in smaller steps. 

\textbf{3) Intractable Modeling.} Even analytically, it quickly becomes intractable to reason about the tens to hundreds of tuning parameters in modern key-value stores and how they interact with all the different mutually exclusive design components to lead to different performance properties. An entirely new performance model is often needed for each permutation of design components, and that the number of possible permutations increases exponentially with respect to the number of components available. Creating such an extensive set of models and trying to optimize across them quickly becomes intractable. This problem gets worse as systems mature and more components get added and it boils down to manual tuning by experts.

\textbf{The Design Continuum Spirit.}
Our work helps with this problem by formalizing this data layout design space so that educated decisions can be made easily and quickly, sometimes even automatically. Design continuums deal with even more knobs than what existing systems expose because they try to capture the fundamental design principles of design which by definition are more fine-grained concepts. For example, a sorted array is already a full data structure that can be synthesized out of many smaller decisions. However, the key is that design continuums know how to navigate those fine-grained concepts and eventually expose to designers a much smaller set of knobs and a way to argue directly at the performance tradeoff level. The key lies in constructing design continuums and key-value store systems via unified models and implementations with continuous data layout design knobs rather than swappable components.

For example, the advantage of supporting LSH-table by continuum expansion rather than as an independent swappable component is that the bridging process adds new intermediate designs into the continuum with appealing trade-offs in-between. The new continuum allows us to gradually transition from a tiered LSM-tree into LSH-table by increasing the size ratio in small increments to optimize more for writes at the expense of range reads and avoid overshooting the optimal performance properties when tuning.

\section{Enhancing Creativity}

Beyond the ability to assume existing designs, a continuum can also assist in identifying new data structure designs that were unknown before, but they are naturally derived from the continuum's design parameters and rules.

\begin{figure}
    \includegraphics[width=.45\textwidth]{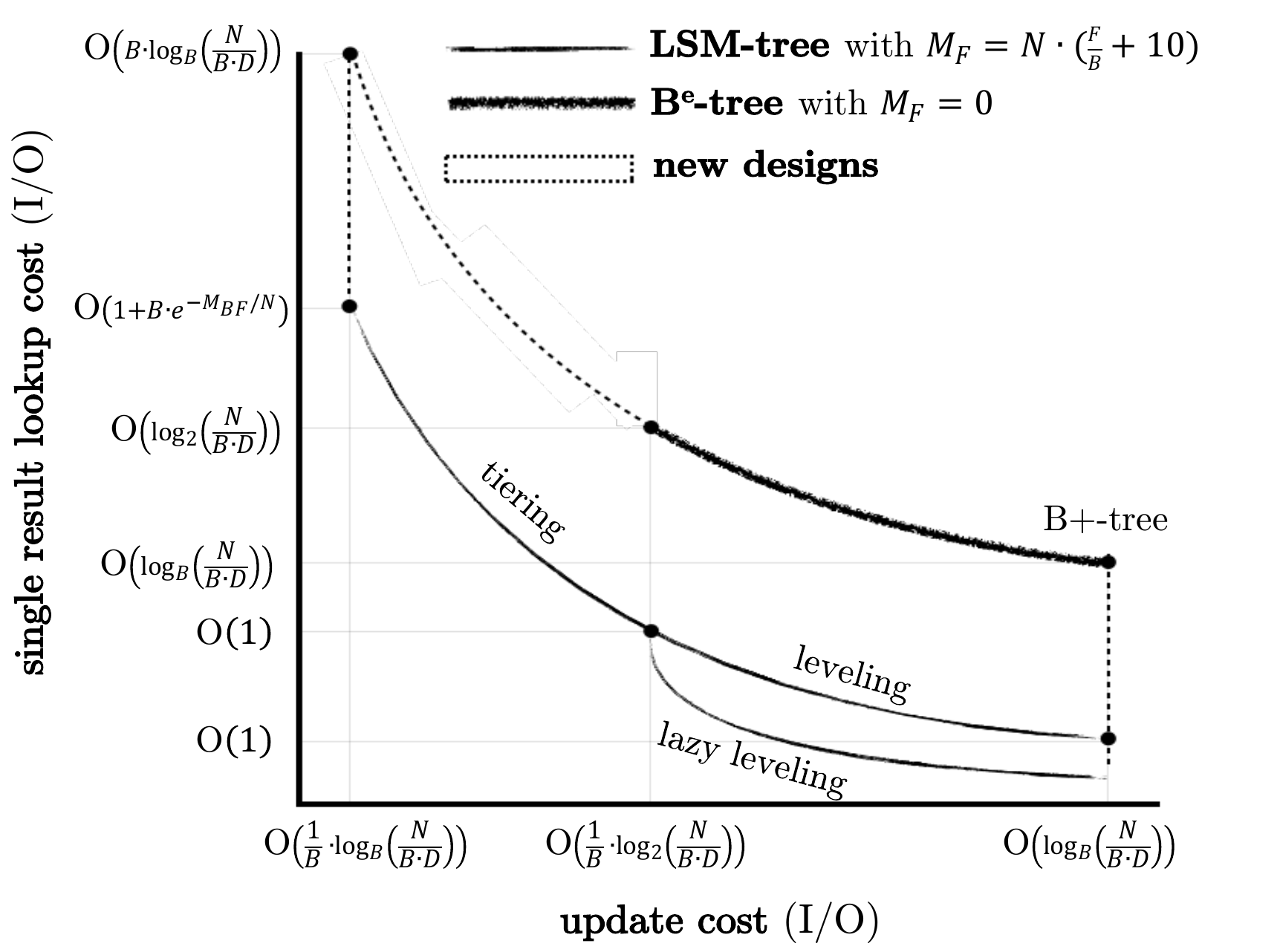}
    \vspace{-1em}
    \caption{Visualizing the performance continuum.}
    \label{f:vcontinuum}
\end{figure}

For example, the design continuum we presented in this paper allows us to synthesize two new subspaces of hybrid designs, which we depict in Figure \ref{f:vcontinuum}. The first new subspace extends the $B^\epsilon$tree curve to be more write-optimized by increasing $Z$ and $K$ to gather multiple linked nodes at a level before merging them. 
The second subspace connects $B^\epsilon$tree with LSM-tree designs, allowing first to optimize for writes and lookups at hot levels by using Bloom filters and fence pointers, and second to minimize memory investment at cold levels by using fractional cascading instead. Thus, we turn the design space into a multi-dimensional space whereon every point maps onto a unique position along a hyperplane of Pareto-optimal performance trade-offs (as opposed to having to choose between drastically different designs only).

In addition, as the knobs in a bijective continuum are dimensions that interact to yield unique designs, expanding any knob's domain or adding new knobs during the bridging process can in fact enrich a continuum with new, good designs that were not a part of the original motivation for expansion. Such examples are present in our expanded continuum where our original goal was to include LSH-table. For example, fixing $K$ and $Z$ to 1 and increasing the size ratio beyond $B$ towards  {\small $\nicefrac{N}{(D \cdot P)}$} allows us to gradually transition from a leveled LSM-tree into a sorted array (as eventually there is only one level). This design was not possible before, and it is beneficial for workloads with many range reads. In this way, the bridging process makes a continuum increasingly rich and powerful.

\textbf{What is a new Data Structure?} There are a number of open questions this work touches on. And some of these questions become even philosophical. For example, if all data structures can be described as combinations of a small set of design principles, then what constitutes a new data structure design? Given the vastness of the design space, we think that the discovery of any combination of design principles that brings new and interesting performance properties classifies as a new data structure. Historically, this has been the factor of recognizing new designs as worthy and interesting even if seemingly ``small'' differences separated them. For example, while an LSM-tree can simply be seen as a sequence of unmerged B-trees, the performance properties it brings are so drastically different that it has become its own category of study and whole systems are built around its basic design principles.

\section{The Path to Self-design}
Knowing which design is the best for a workload opens the opportunity for systems that can adapt on-the-fly. While adaptivity has been studied in several forms including adapting storage to queries \cite{Idreos2007,Alagiannis2014,Arulraj2016,Kennedy2015,Idreos2011,Dittrich2011,Idreos2011a,Liu2016}, the new opportunity is morphing among what is typically considered as fundamentally different designs, e.g., from an LSM-tree to a B$^{+}$tree, which can allow systems to gracefully adapt to a larger array of diverse workload patterns. Design continuums bring such a vision a small step closer because of two reasons: 1) they allow quickly computing the best data structure design (out of a set of possible designs), and 2)~by knowing intermediate data structure designs that can be used as transition points in-between ``distant'' designs (among which it would otherwise be too expensive to transition). 

There are (at least) three challenges on the way to such self-designing systems: a) designing algorithms to physically transition among any two designs, b) automatically materializing the needed code to utilize diverse designs, and c)~resolving fundamental system design knobs beyond layout decisions that are hard to encapsulate in a continuum. Below we briefly touch on these research challenges, and we show hints that they are likely possible to be resolved.     

\SetKwProg{update}{Update}{}{}
\SetKwProg{pointlookup}{PointLookup}{}{}



	



\begin{figure}[t]
  \includegraphics[width=0.5\textwidth]{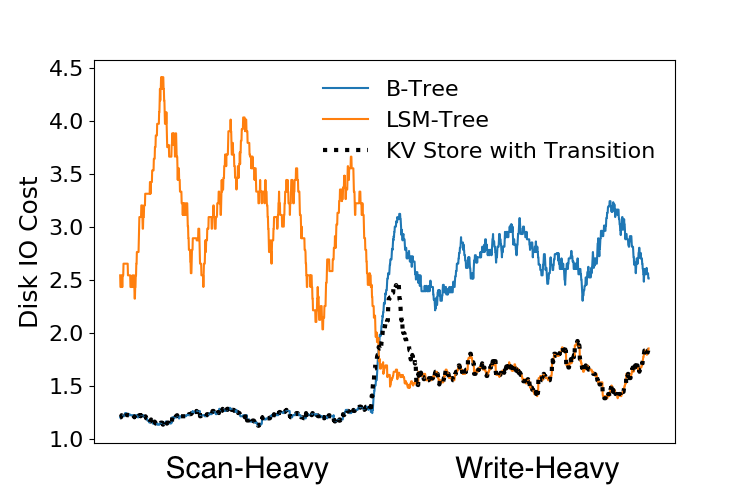}
   \vspace{-2em}
   \caption{Potential benefit of on-the-fly transitions between B$^{+}$tree and LSM-tree.}
   \label{f:transition}
\end{figure}

\textbf{Transitions.}
As in all adaptive studies, we need to consider the cost of a transition. The new challenge here is transitioning among fundamentally different designs. For example, assume a transition between a Leveled LSM-tree and B$^{+}$tree. Even though at a first glance these designs are vastly different, the design continuum helps us see possible efficient transitions; The difference in the specification of each structure on the design continuum indicates what we need to morph from one to the other. Specifically, between an LSM-tree and B$^{+}$tree, merging and fence pointers characterize the main design differences and so the transition policies should depend on these design principles. For example, one way to do such a transition is to wait until the LSM-tree is in a state where all of the entries are at the bottom level and then build the B$^{+}$tree off of that level so that we don't have to copy the data (similar to how we build the internal index of B$^{+}$tree when we do bulk loading). Effectively waiting until merging is not a difference between the source and target design. A second option is to preemptively merge the levels of the LSM-tree into a single level so we can build the B$^{+}$tree off of that level without waiting for a natural merge to arrive. 
A third option is a compromise between the two: we can use the bottom level of the LSM-tree as the leaf layer of the B$^{+}$tree (avoiding copying the data) and then insert entries from the smaller levels of the LSM-tree into the B$^{+}$tree one by one.
We discuss examples of transition algorithms in detail in Appendix \ref{appendix:transition_details}.

The opposite transition, from a B$^{+}$tree to an LSM-tree, is also possible with the reverse problem that the scattered leaves of the B$^{+}$tree need to represent a contiguous run in an LSM-tree. To avoid a full write we can trick virtual memory to see these pages as contiguous \cite{Schuhknecht2016}. The very first time the new LSM-tree does a full merge, the state goes back to physically contiguous runs. 
We explore the implementation-level concerns of keeping the data layouts of such designs interoperable in Appendix \ref{appendix:interoperability}.

Figure \ref{f:transition} depicts the potential of transitions. During the first 2000 queries, the workload is short-range scan heavy and thus favors B$^{+}$tree. During the next 2000 queries, the workload becomes write heavy, favoring LSM-Trees. While  pure LSM-tree and pure B-tree designs fail to achieve globally good performance, when using transitions, we can stay close to the optimal performance across the whole workload. The figure captures the I/O behavior of these data structure designs and the transitions (in number of blocks). 
Overall, it is possible to do transitions at a smaller cost than reading and writing all data even if we transition among fundamentally different structures. 
The future path for the realization of this vision points to a \textbf{transition algebra}.  

\begin{algorithm}[t] 
	\pointlookup{($searchKey$)}{
		\scriptsize
		
		\If{$M_{B} > E$}{ 
			entry := $buffer$.find($searchKey$)\;
			\If{entry}{
				\textbf{return} \textit{entry};
			}
		}
		\tcp{Pointer for direct block access. Set to root.}
		$blockToCheck$ := $levels$[0].runs[0].nodes[0]\;

		\For{$i\gets0$ \KwTo $L$}{
			\tcp{Check each level's runs from recent to oldest.}
			\For{$j\gets0$ \KwTo $levels$[$i$].runs.count}{
				\tcc{Prune search using bloom filters and fences when available.}
				\uIf{$i < (L-Y)$ \tcp{At hot levels.}}{
					$keyCouldExist$ := $filters$[$i$][$j$].checkExists($searchKey$)\;
					\eIf{$!keyCouldExist$}{
						\textbf{continue};
					}{
						$blockToCheck$ := $fences$[$i$][$j$].find($searchKey$)\;
					}
				}

					\tcc{For oldest hot run, and all cold runs, if no entry is returned, then the search continues using a pointer into the next oldest run.}
					$entry$, $blockToCheck$ := blockToCheck.find($searchKey$)\;
					\If{$entry$}{
						\textbf{return} \textit{$entry$};
					}
			}
		}
		\textbf{return} \textit{$keyDoesNotExist$};
	}
	\caption{Lookup algorithm template for any design.}
	\label{algo:pointlookup}
\end{algorithm}

\textbf{Code Generation.}
Tailored storage requires tailored code to get maximum performance \cite{Alagiannis2014}. The continuum provides the means towards such a path; since there exists a unifying model that describes the diverse designs in the continuum, this means that we can write a single generalized algorithm for each operation $o$ that can instantiate the individual algorithm for operation $o$ for each possible designs. For example, Algorithm \ref{algo:pointlookup} depicts such a generalized algorithm for the point lookup operation for the design continuum we presented in this paper.

\begin{figure}[t]
\centering
    \includegraphics[width=.48\textwidth]{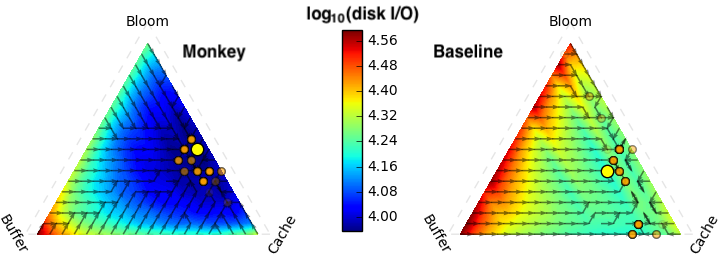}
    \vspace{-2.5em}
    \caption{Navigating memory allocation by learning.}
    \label{f:learning}
\end{figure}

\textbf{Learning to go Beyond the Continuum.} 
We expect that there will likely be critical design knobs that are very hard or even impossible to include in a well-constructed design continuum. 
The path forward is to combine machine learning with the design continuum. 
Machine learning is increasingly used to tune exposed tuning knobs in systems \cite{Aken2017,Pavlo2017}. 
The new opportunity here is the native combination of such techniques with the system design itself. 
For example, consider the critical decision of how much memory resources to allocate to the cache. 
What is hard about this decision is that it interacts in multiple ways with numerous other memory allocations that uniquely characterize a design (specifically the write buffer, the bloom filters, and the fence pointers in our design continuum) but it is also highly sensitive to the workload. 
However, we can use the generalized cost formulas of the continuum to derive formulas for the expected I/O savings if we increase the memory in any memory component. 
We can then use these estimates to implement a discrete form of stochastic gradient descent (SGD). 
Figure \ref{f:learning} shows an example of our results for a skewed workload where we tested two instances of our continuum, the Monkey design \cite{Dayan2017} which optimizes bloom filter allocation and a Leveled LSM-tree design with fixed false positive rates across all bloom filters. 

We evaluate three gradients (pertaining to cache, write buffer and bloom filter budget) at every grid point along the simplex of simulated LSM-trees with constant total memory. We then overlay an arrow on top of the disk access contour plot pointing from the lowest gradient component to the highest gradient component (we move 8 bytes from one component to the other every time). 
Finally, for each grid location, the process follows the arrows until we either reach the edge of the simplex or a previously visited point. 
We then plot partially opaque orange dots at the I/O minima for 3-way memory tunings, predicted by the SGD approach beginning at each simplex starting point.
The set of visible orange dots represents the most frequently predicted I/O minima by the overall SGD process, showing there is general clustering of predicted minima. 
The yellow dot represents a global I/O minimum found experimentally. 
Tests with numerous other workloads also indicate that although as expected the overall optimization problem is sometimes non-convex, we can usually reach a point close to the optimum. 
The details of this methodology for predicting optima for memory tunings using an SGD-based approach are outlined in Appendix \ref{appendix:beyond}.
The net result is that design continuums can be blended with ML approaches to co-design a tailored system that both \textbf{knows} how to navigate a vast space of the design space and \textbf{learns} when needed to navigate design options that are hard to deterministically formulate how they will interact with the rest of the design. 

\section{Next Steps}
\label{sec:summary}


Research on data structures has focused on identifying the fundamentally best performance trade-offs. We envision a complementary line of future research to construct and improve on design continuums. The overarching goal is to flexibly harness our maturing knowledge of data structures to build more robust, diverse and navigable systems. Future steps include the construction of more and larger continuums, and especially the investigation of broader classes of data structure design, including graphs, spatial data, compression, replication as well as crucially more performance metrics such as concurrency, and adaptivity. The most challenging next step is whether the construction of design continuums itself can be (semi-) automated. 

\section{Acknowledgments}
Mark Callaghan has provided the authors with feedback and insights on key-value store design and industry practices repeatedly over the past few years and specifically for this paper. This work is supported by the National Science Foundation under grant IIS-1452595. 


\bibliographystyle{abbrv}

\bibliography{to-be-added,library-short,new}

\appendix{}
\section{Super-Structure Instances} \label{appendix:instances}
We discuss four instantiations of the design continuum in this section.
Lazy-Leveled LSM-tree, Leveled LSM-tree, B$^{\epsilon}$tree, and LSH-table
instances are shown in
Figure \ref{fig:instances}.
while one instance (LSH-table) is expressable in the extension of the design continuum.
In this section, we explain how these four example instantiations are formed.

\subsection{LSH-table}

LSH-table is a structure with a single hot level.
This hot level is the log portion of the structure,
and has growth factor $T = \frac{N \cdot E}{M_B}$,
meaning that the single level log 
can grow up to a capacity sized to fit
up to all of the dataset in the first level of the data structure.
The typical compaction behavior that LSH-table requires as 
its managed dataset grows can be described by the two merge thresholds defined by the
design continuum.
The LSH-table will reach compaction capacity when the number of runs exceeds $T-1$,
indicated by the cold merge threshold $Z = T-1 $,
which specifies that the single hot level will prolong merge for as long as possible.
The hot merge threshold $K$ is only provided
for completeness for single level structures like the LSH-table; 
it is of note to mention that for the design continuum the hot merge threshold parameter does not apply to the
ultimate hot level, rather the ultimate hot level is governed by $Z$ the cold merge threshold.
The LSH-table node size could grow up to $N/B$, depending on how compaction works on the log part.
For instance, a stream of $N$ sequential inserts all compacted at once would compact to a single contiguous 
section of $N/B$ pages which is analagous to a max node size $D$ of
$N/B$ pages.

Two main memory navigation features of LSH-table design differ from 
that of other instantiations of the design continuum: 
first, a hash table is used as a filter and requires on order of one key size ($F$) for each data entry,
and second, to support range scan pruning on every page worth of data (every $B$ entries) in the log,
 a fence of order size $F$ must be maintained in memory.

Note that in the extended design continuum a hash table is allowed to take the place of what was once the Bloom filter budget as defined by the original design continuum.
Implementations of LSH-table which do not need to handle range scans, often omit the fence maintenance feature,
thereby keeping $M_{FP}$ effectively zero, and allocating all of $M_F$ to the hash table.
We illustrate an LSH-table example
in the top left of Figure \ref{fig:instances}; with a single large hash-table and fence pointers enabled.

\subsection{Leveled LSM-tree}
Leveled LSM-tree is a structure with multiple hot levels,
with a growth factor of $T$; each level of the LSM-tree is $T$ times larger in capacity
than the previous level.
All levels but the ultimate hot level, are controlled by the hot merge threshold $K$,
which is set to 1, allowing no more than a single run
at every hot level. The ultimate level is controlled by cold merge threshold $Z$, 
which is also set to 1 for completeness.
Leveling corresponds to maximizing the amount of merge greediness,
bounding the number of runs at each level, and reducing space amplification,
benefiting short and long scans at the expense of
merging updates potentially multiple times at each level.
Figure \ref{fig:instances} bottom left, illustrates a Leveled LSM-tree
that uses fence and filter memory budget $M_F$ amounting to $\frac{F}{B}$ bits 
per entry for fences and 10 bits per entry for Bloom filter.

\subsection{Lazy-Leveled LSM-tree}
Lazy-Leveled LSM-tree is a structure with multiple hot levels,
with a growth factor of $T$; each level of the LSM-tree is $T$ times larger in capacity
than the previous level.
All levels but the ultimate hot level are controlled by the hot merge threshold $K$,
which is set to $T-1$, allowing up to $T-1$ runs to build up
at each of these levels
and maximally relaxes the number of times merge occurs
(in comparison to a Leveled LSM-tree).
The ultimate hot level is controlled by the cold merge threshold $Z$,
and only has one large run,
thereby bounding space amplification caused 
from relaxing hot merge threshold $K$ at the first $L-1$ levels of the tree.
Figure \ref{fig:instances} top right illustrates a Lazy-Leveled LSM-tree
that uses fence and filter memory budget $M_F$ amounting to $\frac{F}{B}$ bits 
per entry for fences and 10 bits per entry for Bloom filter.

\subsection{B$^{\epsilon}$tree}
B$^{\epsilon}$tree is a structure with only one hot level and the rest cold levels.
The single hot level is analagous to the first level of a tree below the root node.
The in-memory fence pointers only exist for the first level.
The fence pointers and write buffer are analgous to the child pointers and contents
typically held in a B-tree's root node.
From a performance standpoint, the B$^{\epsilon}$tree instantiation
behaves as if it has a root node \textit{pinned} in memory.
The tree itself has a fanout described by the growth factor $T$, which varies dynamically between 2 to $B$
depending on how full the nodes are.
The hot merge threshold $K$ is defined for completeness, the cold merge threshold $Z$ governs all levels 
(including the single hot level) and is set at 1, describing that each level of the instantiation
is one run. This allows point lookups to navigate the structure only by examining one node at each level,
and following the embedded fence pointers to reach cold levels of the tree.
The batch update mechanism is cascading, as each level becomes full, the entries at that level must be flushed and merged with the overlapping nodes of the 
next largest level, while the embedded fences of each level that point to the next are updated
to reflect the merges that have taken place.
By setting the maximum node size $D$ to 1, the instantiation describes the smallest unit of merge
as a block size.
Figure \ref{fig:instances} bottom right illustrates a B$^{\epsilon}$tree instantiation.

\begin{figure*} 
    \includegraphics[width=\textwidth]{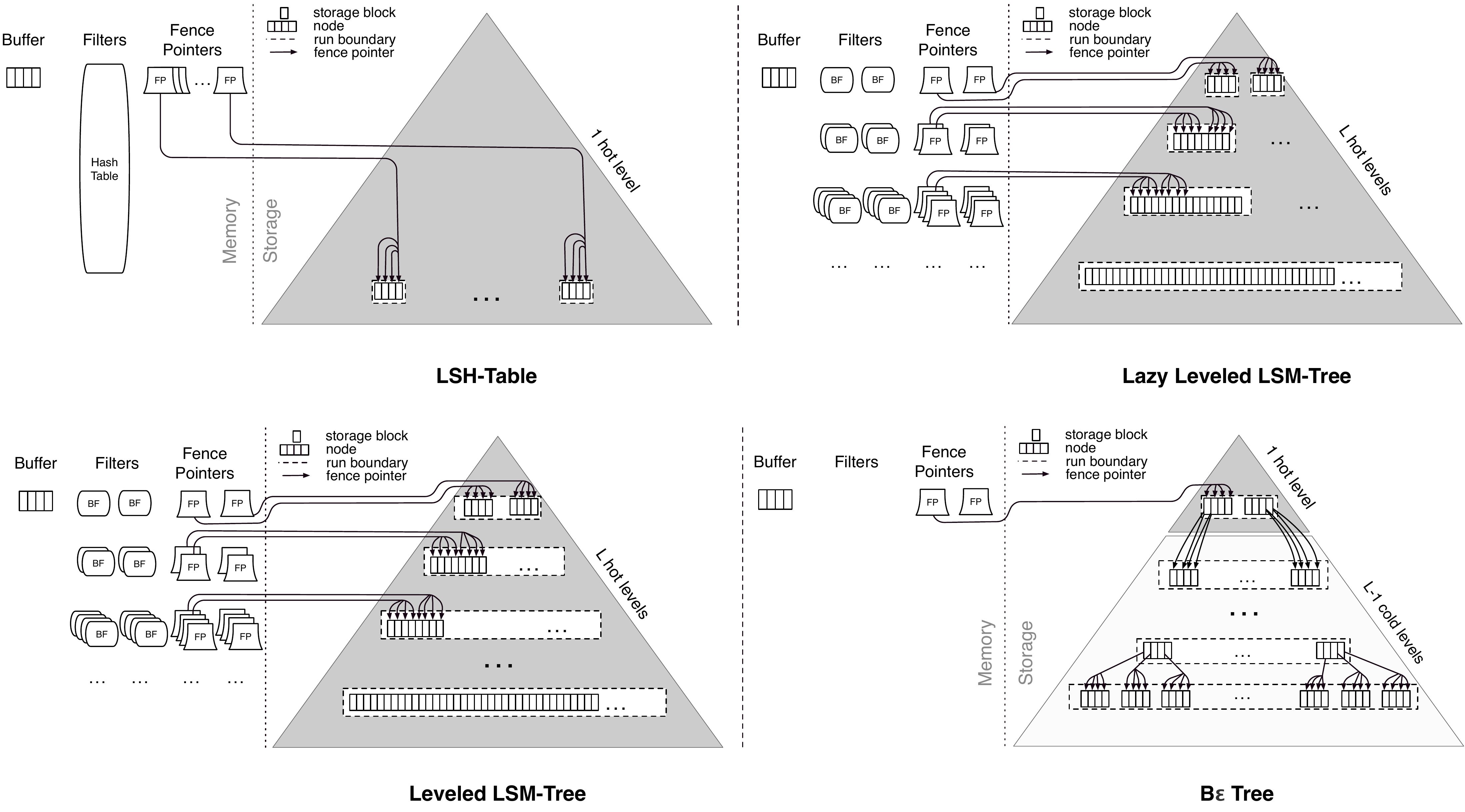}
    \vspace{-2em}
    \caption{Various instantiations of the super-structure from the design continuum.}
    \label{fig:instances}
\end{figure*}

\section{Design Transitions}
In this section we discuss in further detail a preliminary exploration of what is required for facilitating online transitions 
between data structure designs on the design continuum.

\subsection{Transition Algorithms} \label{appendix:transition_details}
As an example we outline possible transition algorithms between an LSM-tree to a B$^{+}$tree. 
We give two different strategies and then we discuss how to chose between them and how to holistically design and schedule transitions. 

\textbf{Sort-Merge Transition.}
In an LSM-tree, the data is organized into several sorted runs with some obsolete entries that result in space amplification.  In order to condense this data into the leaf level of a B$^{+}$tree, we must merge all of these runs into a dense, sorted array that becomes the leaf level.  In this approach, we force a sort-merge of all the runs in the leveled LSM-tree structure. This has a performance cost of reading in all the runs' pages, and then writing each of the sorted pages to disk.  This method has optimal space amplification properties, since the resulting array is compact. Moreover, since we would usually switch to a B$^{+}$tree when the workload is more read-heavy than write-heavy, a compact array would be better (fuller B$^{+}$tree leaves and therefore fewer disk pages of data to scan).

\textbf{Batch Inserts Transition.}
In some cases (e.g. when there is relatively little data in the levels above the bottom of the LSM-tree), we may be able to avoid reading and writing most of the data.  To achieve a more lightweight transition in these cases, we can take the bottommost (largest) run of the LSM-tree and treat it as the leaf level of a B$^{+}$tree.  We then insert the entries from the upper levels (levels above the bottom) of the LSM-tree into this B$^{+}$tree. By performing these inserts in sorted order, we can ``batch insert'' the entries into the B$^{+}$tree. In particular, consider a node $N$ of the leaf layer for which there are entries in the upper levels to be inserted into $N$. We load that node into memory. Then, we iterate through the elements in the upper levels corresponding to the range of $N$ - call this set of elements $R$. As we do so, we merge the elements of $N$ and $R$, making new leaf-level nodes in memory. These new nodes are flushed to disk as they are made. As an example, if the set of elements $R$ to be inserted into node $N$ consists of just one element, then the above process is equivalent to a B$^{+}$tree node split.

\begin{figure}
    \vspace{-20pt}
    \centering
    \raisebox{-0.5\height}{\includegraphics[width=0.35\textwidth]{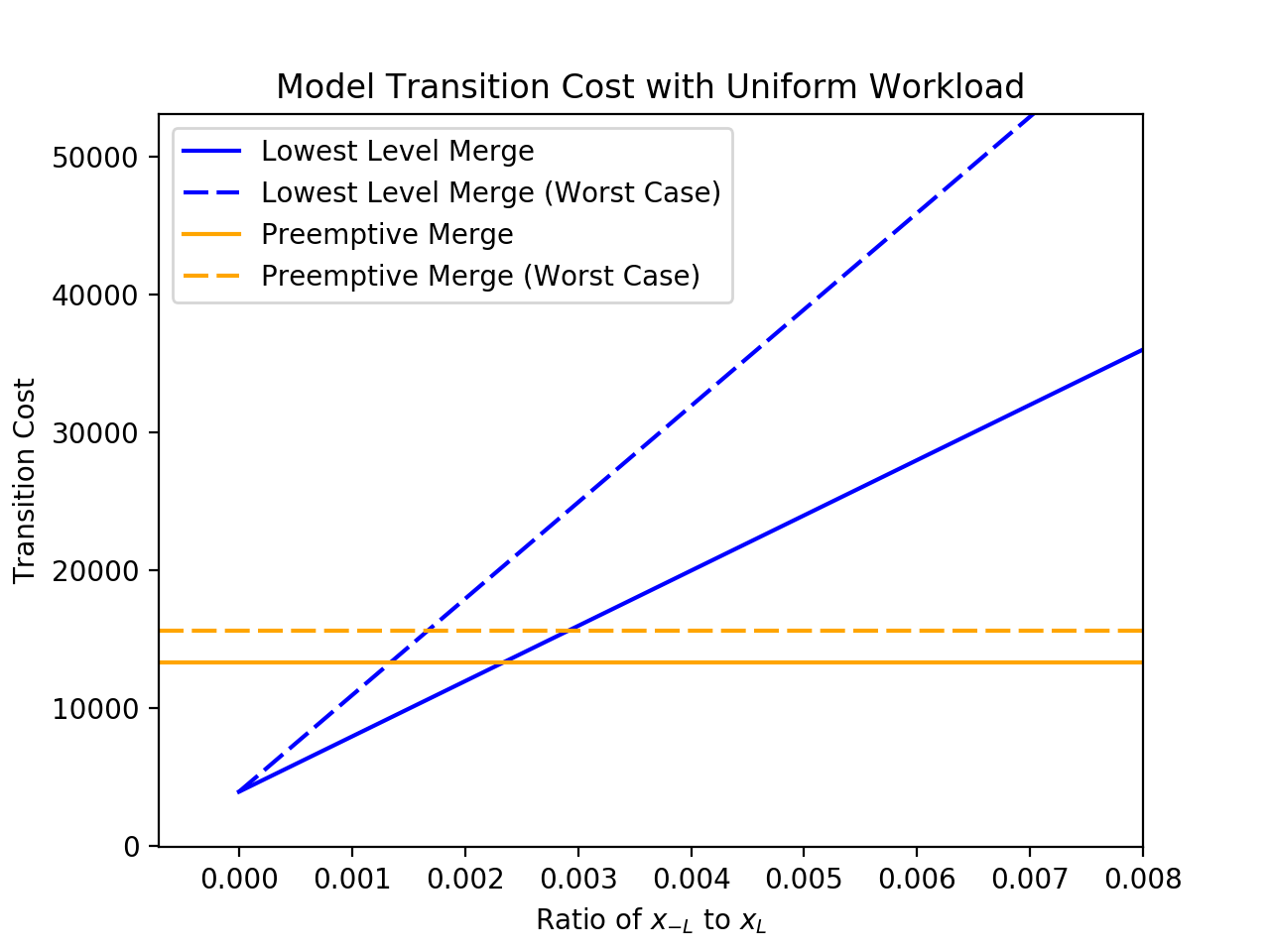}}
    \hspace*{-0.45cm}
    \raisebox{-0.5\height}{\includegraphics[width=0.14\textwidth]{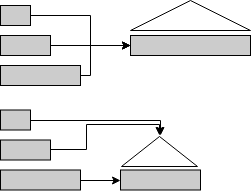}}
    \vspace{-10pt}
    \caption{Tradeoff Between Merge Sort (Upper Right) and Batch Insert (Lower Right)}
    \label{fig:my_label}
\end{figure}

\textbf{Choosing the Optimal Transition.}
Compared to the sort-merge approach, the batch inserts method may be more efficient because we may not insert entries into every node of the B$^{+}$tree, saving reading and writing costs of certain nodes. On the other hand, it has worse space amplification properties because some pages may be partially empty; this can result in more disk IO. We show that, in both average and worst case analysis, the batch inserts approach supersedes the sort-merge approach when the size of the upper runs is below a certain threshold. To illustrate this point, we present the results for a simple worst case analysis of a basic LSM-tree with node of size $1$ page. 

Assume an LSM-tree with $L$ levels where $x_i$ is number of entries in the $i$th level and $d$ is the size of each entry in bytes and a machine where $p$ is the page size in bytes and $\phi$ is the ratio of a disk write to a disk read.  The cost of a sort merge transition is $C_{SM} = \left( \sum_{i=1}^{L} \left\lceil \frac{d \cdot x_i}{p} \right\rceil \right) (1 + \phi)$.  The worst-case cost of a batch insert transition is $C_{BI} = \left\lceil \frac{d \cdot x_L}{p} \right\rceil + \sum_{i=1}^{L - 1} x_i \left(\frac{d}{p} + 1 + 2\phi \right)$.  Let $x_{-L}$ be the number of entries in all levels above $L$; we choose to do a batch insert transition only when $C_{BI} < C_{SM}$, in other terms when $\frac{x_{-L}}{x_L} < \frac{d \phi}{p + (2p - d) \phi}$. See Figure $2$ as a visualization of this tradeoff: Using our cost model, we show that as the ratio $\frac{x_{-L}}{x_{L}}$ decreases, the cost of the batch insert method decreases while the sort-merge method is unaffected.  This is because as $x_L$ becomes relatively larger, the LSM-tree is more "bottom-heavy," and we can potentially use fewer disk operations when executing a batch insert transition.  In this case, $\frac{x_{-L}}{x_{L}} = 0.2$ is our threshold point, below which it is cheaper to use a batch insert transition.

\textbf{Gradual Transition.}
In many situations a sudden spike in query latency would be undesirable. Transitions should be lightweight but also we should be able to control the transition cost with respect to the current system status.  We propose a robust, gradual transition algorithm that allows us to transition from an LSM-tree to a B$^{+}$tree over the course of an arbitrary number of steps. This algorithm exhibits a tradeoff between total transition cost and query latency during the transition phase. At each step, we take the $k$ smallest pages of data from the LSM-tree and insert these pages into the B$^{+}$tree at the leaf level, updating the intermediate in-memory nodes accordingly.  We keep track of a "threshold value" $\theta$, which is the highest key value from our LSM-tree that we have inserted into our B$^{+}$tree.  Thus, each step will only require us to append $k$ pages to the leaf level of the B$^{+}$tree since we know that the key values contained in these pages will all be greater than $\theta$ at the previous step of this algorithm.  After each step we update $\theta$.  We sort-merge the first $k$ pages' worth of data to generate the values to add to our B$^{+}$tree, and we can do this efficiently by generating in memory one page at a time and then writing that page to disk.  This allows us to only have to read and write to disk each entry in the LSM-tree once.  To handle reads and writes while the transition is underway, we check if the query key is greater than or less than $\theta$, and this tells us whether to execute the query on the LSM-tree, the B$^{+}$tree, or possibly both in the case of range queries. The total cost of the transition will include the cost of reading and writing data in the original LSM tree, the cost of reading and writing any new data added to the LSM tree during the transition, and the opportunity cost of queries during the transition.



\textbf{B$^{+}$tree to LSM-tree Transitions.} A naive transition from a B$^{+}$tree to an LSM-tree could take one of two approaches: we could insert the leftmost $k$ pages of our B$^{+}$tree into our LSM-tree over $k$ steps. 

The second approach, which we detail further in the proceeding sections, would allow us to treat the B$^{+}$tree as the bottom-most level of the LSM tree without reading its contents from disk. The naive approach to this would requires re-writing all the non-contiguous leaves of the B$^{+}$tree into a contiguous block of memory, since B$^{+}$trees leaves are not guaranteed to be contiguous while LSM runs \textbf{are} traditionally required to be contiguous. This operation would depend on a variable $x$: The "degree of fragmentation" given by the ratio of the number of contiguous chunks in the data layer to $NE$, the total bits of data in the data layer. The cost of this naive transition would be linear in the number of pages to read and write, namely $\alpha NEx + \alpha NE(1-x) + NE$, where $\alpha$ is an amplification factor to account for the amount of empty space in the B$^{+}$tree leaves (since B$^{+}$tree leaves are usually not completely full; i.e. $\frac{\alpha N}{B}$ pages will contain $N$ data entries).

This presents a clear opportunity for optimization. Since the B$^{+}$tree's leaves are already internally sorted and in sorted order, the data would simply be copied over and no sorted step takes place, unlike in the LSM -> B$^{+}$tree transition, where rewrites were \textbf{necessary}. 

\subsection{Supporting Data Layout Interoperability} \label{appendix:interoperability}
As a self-designing system undergoes design transition, the architecture must employ techniques to support seamless
interoperabilty between data layouts as the transition is underway; it is crucial that as many components of the data layout
from the previous design can be reused in the new design to save work for the system. 
Here we propose an implementation-level abstraction for storage layer management with the objective to mitigate the cost of data rewrite during a B$^{+}$tree to LSM-tree transition.

\textbf{Mapping Page IDs to Disk Locations.} To avoid simply copying the data into a contiguous memory region, we propose an abstraction layer to allow LSM trees to work with non-contiguously stored runs and levels. This abstraction layer would ``deceive'' the LSM tree into believing that it's using contiguous memory regions, and so would require no change to the logic of the LSM tree, and would avoid rewriting any disk data. Instead, only memory indexing structures would change.  



This abstraction would cause the LSM tree to manage its references to data in terms of page IDs instead of in terms of addresses to disk pages. LSM trees require contiguously stored runs due to their fence pointer architecture. Each run of the LSM tree is indexed into via a set of fences that indicate which value ranges are stored in which pages. In order to be space efficient, this indexing system stores only one pointer (the pointer to the first page of data in the run), has one "fence" for each page which indicates the largest value stored in that page. These fences are iterated through to find the poge that contains the value(s) of interest, and the appropriate page is then accessed \textbf{using the single pointer and an offset.} It is this single-pointer and offset mechanism that would break down if the pages that make up a run of an LSM tree were not stored contiguously. Page IDs solve this problem by providing a mapping from contiguous numbers (page IDs) to actual disk page addresses, which no longer need to be contiguous.

Therefore, the LSM tree's implementation would sense little difference between referencing disk regions and page IDs. The page IDs would be redirected through our abstraction layer to the appropriate disk pages.

Since the abstraction is only useful when a B$^{+}$tree is being converted to an LSM tree, and we expect all the B$^{+}$tree's data to reside in the bottom layer of the newly formed LSM tree, \textit{our abstraction is only necessary for the last level of the LSM tree, until the first merge occurs.} As the LSM tree is updated, upper levels can be stored contiguously on physical pages, which will be a faster writing approach anyway. 

Additionally, once the time for the first merge operation arises, we will merge the penultimate layer with the last layer (the one behind the abstraction) and write the new last layer to disk contiguously. Since the last layer is now contiguous, our data structure is not necessary.

\textbf{Lightweight Indirection Mapping.}
The most obvious approach to map page IDs to disk locations is to use a hash table that stores page IDs as keys and disk locations as values. 

This structure, however, would require memory on the order of $O(N)$. Additionally, a one-to-one mapping would not be necessary for the majority of pages in our data layer since many of the pages that comprise the data layer may be stored contiguously, since if we convert an LSM Tree to a B$^{+}$tree, then back, we observe that the only case where the contiguous runs of the LSM tree are broken is in places where new data pages are inserted. If we bound the number of insertions while in the B$^{+}$tree data structure, we can maintain fairly contiguous regions of data. We can therefore define a compression scheme for our mapping data structure. Instead of mapping page IDs to disk locations, we will map data "regions" to disk locations.
For instance, if page IDs 2 to 1002 are stored contiguously, instead of being represented by 1000 mappings from page IDs to physical pages, the region would be represented by a single entry containing the starting page ID and the disk location of the first page. The rest of the pages may be accessed using offset arithmetic. 
This data structure would require memory on the order of $O(xN)$, where $x$ is the fragmentation percentage of the data layer.

Therefore, in order to redirect page IDs to disk locations, we will make use of a miniature B$^{+}$tree like structure. Since the keys on a B$^{+}$tree are sorted, we can easily search the B$^{+}$tree for a key that is not currently mapped by the B$^{+}$tree, and access (with an offset) he value associated with the largest key less than it. Hence, rather than storing a mapping for every mapped page, we can simply store the beginning Page ID for a run of contiguous pages as our key, and the corresponding beginning pointer for our physical page as the value.

\textbf{Side-Effects for LSM Tree Performance.}
In considering the performance implications of using this abstraction, the key insight is that we must now complicate our earlier algorithms, which costed in terms of disk I/Os, and instead cost in terms of random disk I/Os and contiguous disk I/Os.

We introduce three new variables: 

\begin{itemize}
    \item $c$: Average cost of continuous read
    \item $r$: Average cost of random read
    \item $w$: Average cost of continuous write
\end{itemize}

We arrive at the following conclusions about the use of our storage/compute abstraction:
\begin{itemize}
    \item Use of our API causes 100\% reduction in the disk I/O cost of transitioning from a B$^{+}$tree to an LSM Tree, since our new transition only requires memory operations, but no disk I/Os.
    \item Until the first merge occurs, range queries take significantly longer than they would in a regular LSM tree that does not use our API. Specifically, the degradation in performance is equal to
    \begin{equation}\label{range-loss}
        \alpha NEs(xr + xc - c) + \frac{scNE(T+1)}{(T-1)}
    \end{equation} 
    \item The cost for the first merge differs between our API and the standard algorithm by 
    \begin{equation}\label{fmerge-loss}
        x(NEr-c)
    \end{equation}
     This term is positive since $r>c$ and $N\geq1$.
\end{itemize}

The decrease in performance for a range query in an LSM tree is not a serious concern for our intended use case; since our hybrid data structure would only switch to a B$^{+}$tree in the case of a write-heavy workload, in most cases, the additional cost of the few range queries that occur prior to the first merge will not outweigh the benefits of the faster transition.

\section{Beyond Design Continuums} \label{appendix:beyond}



Bringing additional knobs into the fold with an existing design continuum
adds additional challenges for reasoning about the performance tradeoffs of designs.

\textbf{Complex parameter spaces:} In many cases the knobs might display a high
level of sensitivity to the values of other knobs. This kind of complex
interaction can make the manual construction of a proper design continuum
intractable.

Consider the critical design decision of how much memory to allocate to a cache
in a LSM-tree. Since we have a fixed amount of main memory budget, the cache
budget must interact in multiple ways with other memory
allocation budgets.

In order to increase the size of the cache, 
we must strategize the best way to accordingly distribute the
reduction in budget between the two other components that consume main memory budget: the write buffer and Bloom filter.
For instance, if we choose to take away from the size of the write
buffer, we might assume that more queries will end up touching more levels of the LSM-tree. 
Is this penalty offset by having a larger cache? This
depends on the hit rate of the cache, which ultimately depends on the query
workload.

If we have a workload consisting of a small working set of keys dominating the majority of
queries, then perhaps the resulting improvement from cache hits will
produce an overall improvement in performance.
Alternatively, if we have a workload where the hot working set of keys is larger than
the cache or even main memory, then we might find that the strategy
of maximizing cache size turns into a bad one - e.g. a relatively small proportion of queries
that cause a cache miss and hit disk might be enough to sink overall
performance because by reducing the space allocated to Bloom filters we've
increased their inaccuracy with higher false positive rate (FPRs), 
meaning that each run is more likely to incur the \textit{full} cost of a random I/O probe,
rather than a lower \textit{protected} cost from effective Bloom filters.

If we had instead decided in the beginning to remove more memory from the Bloom
filters rather than from the write buffer, we'd encounter a different set of dilemmas in
the decision making process. On top of this, would increasing the size of the
write buffer be a good decision in the first place? We would need yet another set of
criteria to be able to make that determination.

\textbf{Discretizing parameter spaces:} We might have design knobs that are
not smooth and continuous, but are made of many discrete options.
Moreover, even continuous design knobs
may need to be discretized in practice, to be navigable.

\textbf{Complex workloads:} In real world scenarios we might e.g. find that the
hot working set of keys is not static but changes over time, affecting performance with
it \cite{Xu2014}.

For only a few parameters, we can end up with a situation where even hand-wavy
intuition on tradeoffs becomes complicated. Knowing the exact parameters or the
exact tipping points in which certain design strategies become better or worse is yet
harder. At each turn, we encounter more scenarios where the prevailing strategy is highly
dependent on the co-interaction of other design knobs and the nuances of the workload.

\subsection{A Solution: Stochastic Gradient Descent}

We see a path forward by combining machine learning ideas with design continuum
knowledge to create solutions that approach the optimal for a broader
set of design decisions. The problem, to us, seems analogous to training a
machine learning model:

\textbf{The cost function} that we're minimizing is the data access cost
expressed in random I/O's to storage,
derived from the design continuum.

\textbf{The goal} is to minimize the cost function, i.e. find the parameters
that provide the lowest I/O cost for the given set of queries.

\textbf{The parameters} are the cache size, the write buffer size and the bloom
filter size.

\textbf{The parameter space} is only a subset of 3D space, because total memory
is constrained. For example, if we wanted to increase both the cache size and
the write buffer size we would have to decrease the bloom filter size.

\textbf{The gradient functions} are the estimated I/O savings if we increase
the memory by N bits for any single component. Following the gradients, one
would expect to end up in a minima. While deriving these analytically is
difficult, we come up with reasonable estimates using the design continuum cost
formulas.

\textbf {The dataset} we're training the model on is a key-value workload, i.e. a set of 
queries. We automatically generate ours from probabilistic models, but one
could also use traces of real systems, or perhaps a combination of both.

\subsection{Stochastic Workloads}

To avoid overfitting to a particular set of queries, we define some workload
classes as configurable probabilistic models. We can then generate an ordered
sequence of point reads or updates (i.e. a workload) from any of these. We include
more complex, time-varying workloads (inspired by
\cite{Xu2014} and \cite{Armstrong2013}) that attempt to mimic
realistic settings.

For all the workloads, when we draw a particular key for the first time we
will insert it into the database as a write, and subsequently we will either
look it up or update it with probability $w$.

\textbf{Uniform} queries will be drawn uniformly from a set of $K$ keys. This is
often one in which the cache is unhelpful, but in practice may be unrealistic.
Nevertheless, this is the scenario that many analyses assume.

\textbf{Round-Robin} queries are drawn deterministically using $k_i = (i \mod
K)$, i.e. we iteratively draw each key in sequence, then repeat.
This represents another pathological scenario for a cache: a key that has been
recently written or read is actually a contraindication we will access it
again.

\textbf{80-20} queries (considered in \cite{Dayan2017}) are drawn such that 20\%
of the most recently inserted keys constitute 80\% of the lookups. This is a
simple model to observe the effects of skew.

\textbf{Zipf} queries are distributed according to a zeta distribution, with a
parameter $s$ which describes the skewness. Zipf-distributed queries are
considered in \cite{Leis2013} as another simple proxy for realistically skewed
queries.

\textbf{Discover-Decay} queries are distributed according to the following
stochastic process, inspired by the Chinese Restaurant process \cite{Aldous1985} but
with time decay: with every passing time step, we draw a number of reads $n_r$,
writes $n_w$, and updates $n_u$ assuming queries arrive according to Poisson
processes with configurable rates: \[
\begin{split}
  n_r & \sim \textrm{Pois}(\lambda_r) \\
  n_w & \sim \textrm{Pois}(\lambda_w) \\
  n_u & \sim \textrm{Pois}(\lambda_u)
\end{split}
\]
Once we've drawn our $n_w$ new keys $k_i$, we assign them an initial popularity
$$
\theta_{i} \sim \textrm{Beta}(a_\theta,b_\theta)
$$
\noindent with a random decay rate
$$
\gamma_i \sim \textrm{Beta}(a_\gamma,b_\gamma),
$$
which is the factor by which they exponentially decay each subsequent time
step. At any time $t$, the popularity of each key is given by $p(k_i,t) \propto
\theta_i\gamma_i^{t-t_i}$, where $t_i$ is when the key was inserted. We use
these time-dependent popularities to draw each of our $n_r$ reads and $n_u$
updates from $\textrm{Mult}(\{p(k_i,t)\})$.

\textbf{Periodic Decay} workloads are a simple modification of the
Discover-Decay model where $p(k_i,t)$ now depends not only on the decay rate
$\gamma_i$ but also on a periodic function of the key's age $t-t_i$.  To mimic
the combination of exponential decay and sharp periodic peaking we see in
\cite{Xu2014}, we multiply $\theta_i\gamma_i^{t-t_i}$ by an
inverse cycloid function with period $T$, clamped from 0 to 1, and taken to a
configurable power (to make the cusps sharper or duller) that we call the
cycloid's \texttt{cuspity}.

\begin{figure*}[!]
\centering
\includegraphics[width=0.80\textwidth]{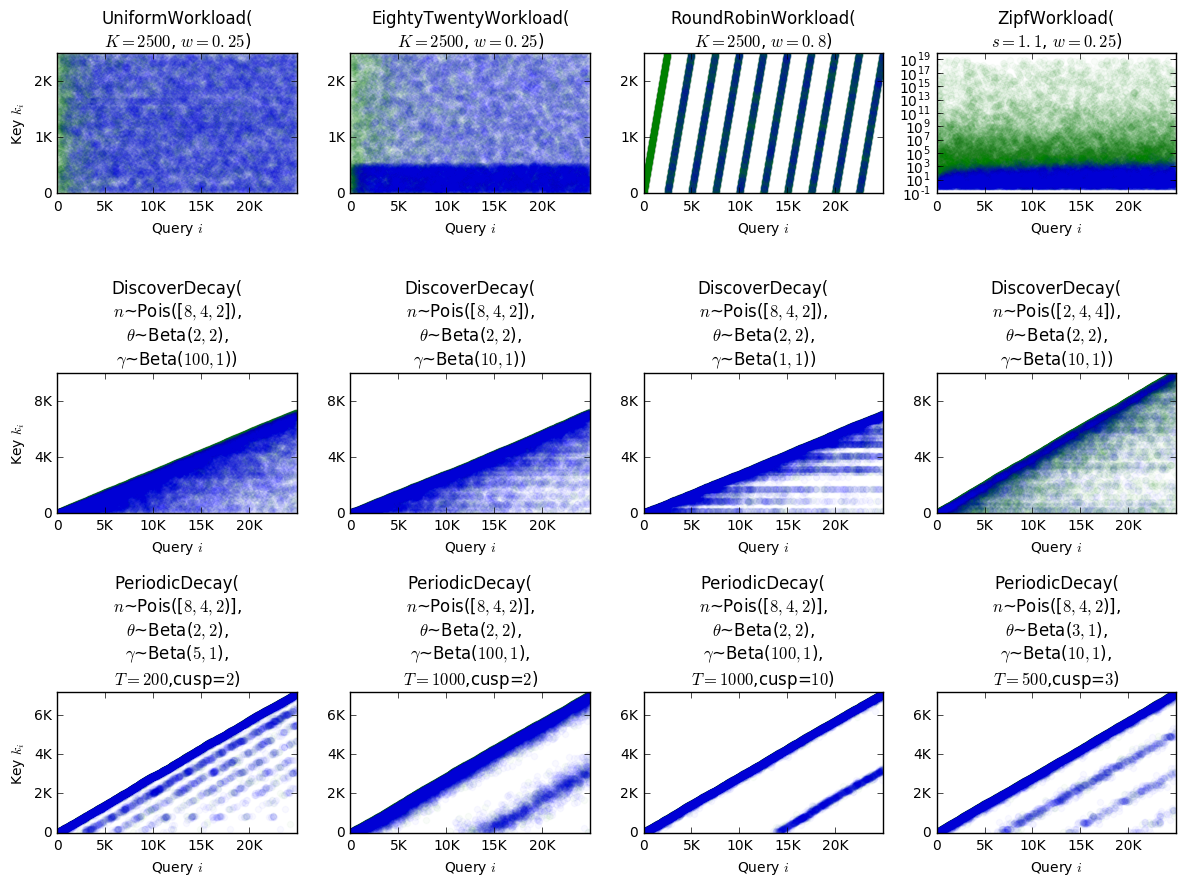}
\caption{Diverse set of workloads used for benchmarking.}
\label{fig:workloads}
\end{figure*}

Examples of these workloads are illustrated in Figure \ref{fig:workloads}.
The first row contains simple workloads where the distribution of key popularities does not
change over time, and where the read/write ratio is a uniform probability.
The second row contains Discover-Decay workloads, consisting of operations
that insert/read/update keys
according to Poisson processes and simulate popularity decays over time. The
third row is a modified version of Discover-Decay that adds a periodic signal
to the decaying popularity with a configurable period and cusp sharpness. Blue
dots represent reads and green dots represent writes (inserts or updates).

\subsection{Modeling}

We derive equations for the number of storage accesses saved by adding N bits more
space to a component. We store simple O(1) space statistics (e.g. number of
queries, number of accesses for each bloom filter, number of times a key was
found in a level) to get easy to compute and reasonable estimates for these.
These loosely follow the general form:

\begin{align*}
    &\Delta accesses = \widehat{hit\_rate}\cdot \Delta{entries} \cdot\widehat{miss\_cost}
\end{align*}

More specifically:

\begin{align*}
    &\Delta cache = \widehat{last\_slot\_hits} \cdot \Delta{entries} \cdot  \widehat{avg\_cost\_of\_miss} \\
    &\Delta buffer = \sum_{l\in L} \widehat{accesses_{l}} \cdot \frac{dM \cdot {level\_ratio}}{Mlevel_{l}} \cdot \widehat{avg\_cost\_of\_miss_{l}}\\
    &\Delta bloom = \sum_{l\in L} \widehat{accesses_{l}} \cdot alloc(l, m+dM) \cdot \widehat{\Delta false\_pos\_rate_{l}}\\
\end{align*}

For full derivations, refer to Section \ref{modeling}.

\subsection{Gradient Descent}

\begin{figure}[!htb]
    \begin{center}
    \includegraphics[width=0.5\textwidth]{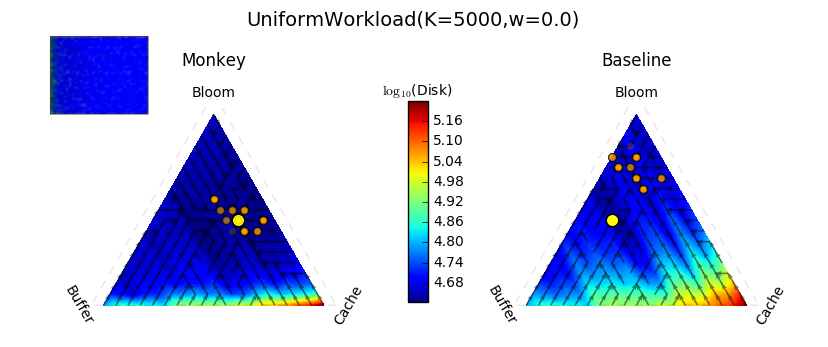}
    \includegraphics[width=0.5\textwidth]{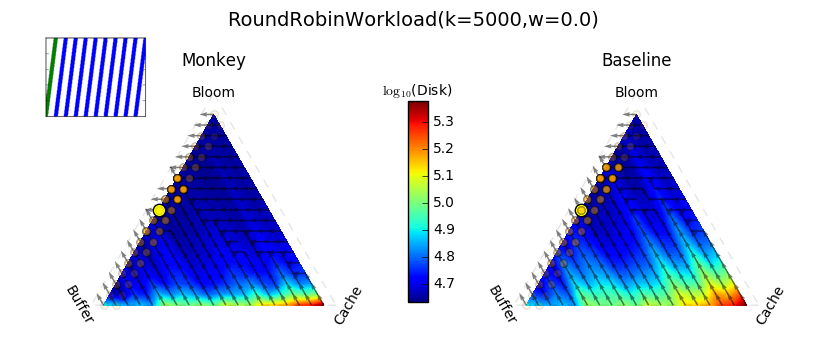}
    \includegraphics[width=0.5\textwidth]{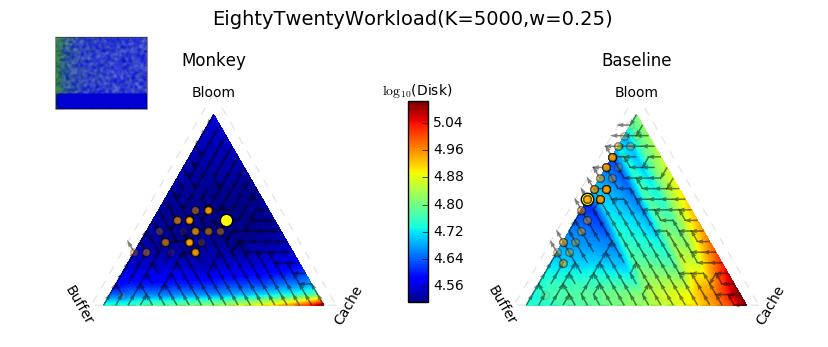}
    \end{center}
\caption{Uniform and Round-Robin simulation results overlaid with gradient
    estimates.}
\label{fig:basicquiv}
\end{figure}

Results for basic workloads can be seen in Figure \ref{fig:basicquiv}. To
display our results, we introduce a specific plot that shows the workload, our
estimated gradients at each point, our estimated best configurations, the
actual performance across the whole parameter space and the actual optimal
configuration.

\textbf{The triangle} shows the parameter space. Each coordinate within the
space represents an LSM-tree with a particular combination of these parameters.
Perpendicular distance from the edge opposite a corner represents the value for
that corner's parameter, e.g. a coordinate in the very center represents an
equal allocation of all three parameters and a coordinate at a corner
represents allocating everything to only that parameter.

\textbf{The arrows} represent our estimated gradient around that coordinate
i.e. given the current set of parameters, which direction our I/O savings model
says you should move to get a better total I/O cost.

\textbf{The orange dots} signify the parameters we predict will have the
minimum I/O cost.  One can start the gradient descent process from any given
initial point. We test our method with every possible initial point to ensure
that it consistently yields good results, which is why there is more than one
prediction, and why the dots have varying opacity of orange color.
Darker orange points indicate parameter configurations
that were more frequently the resulting predictions,
while fainter orange points were less frequently the resulting predictions.

\textbf{The blue-red shading} represents actual I/O cost for each parameter
combination, generated by exhaustively running queries from the workload on an
LSM-tree simulator.

\textbf{The yellow dot} represents the lowest actual minimum I/O cost
determined experimentally, and therefore the optimal allocation of parameters -
our target.

The gradient descent process works as follows:

\begin{itemize}
\setlength{\itemsep}{0pt}
\setlength{\parskip}{0pt}
\setlength{\parsep}{0pt}
\item Ignoring the shading, start at any random initial parameters. Follow the
    arrows until you either hit the edge of the parameter space or hit a point
        where there are no outward arrows from your current location.
\item To evaluate our results, one can look at the plot and should check to see
    that the predicted minima (orange dots) are nearby the actual minimum
        (yellow dot), or failing that, on a grid coordinate with similar IO cost
        (judged by the blue-red shading).
\end{itemize}

Results for basic workloads can be seen in Figure \ref{fig:basicquiv}. For each
workload, we provide results for both the even Bloom filter allocation and the Monkey Bloom filter
allocation schemes.

The uniform workload provides a baseline workload to compare other results to.
The round-robin workload provides an example of a canonical workload that
thrashes the cache to the point where it is useless, and indeed our
recommendation is to allocate no space to the cache. Discontinuities in the
number of levels as we vary the buffer size makes the optimization
non-convex, but Monkey improves both absolute performance and convexity.

\begin{figure}[!htb]
    \begin{center}
    \includegraphics[width=0.5\textwidth]{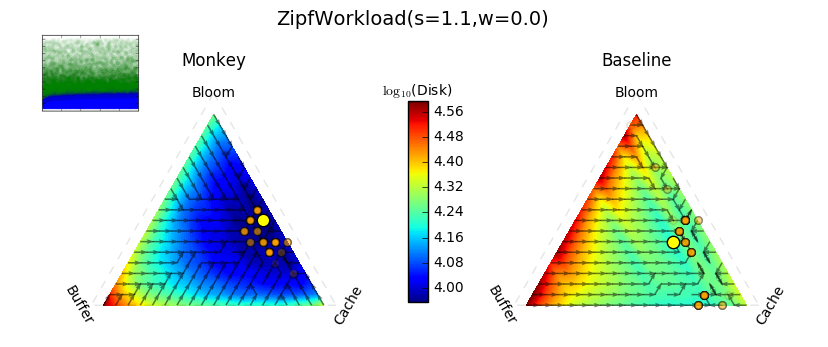}
    \includegraphics[width=0.5\textwidth]{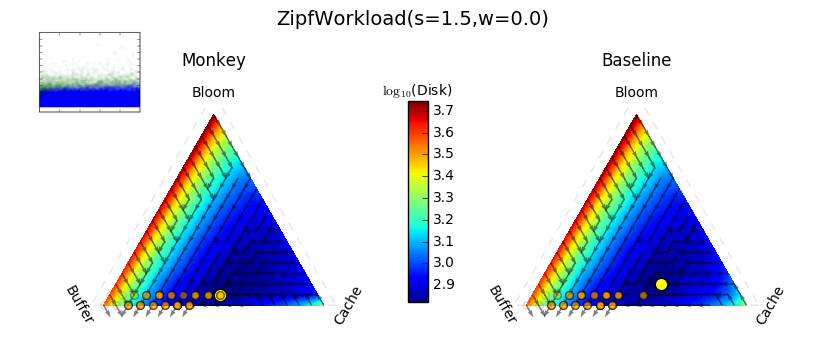}
    \end{center}
\caption{Zipf simulation results overlaid with gradient estimates for $s=1.1$
(lightly skewed) and $s=1.5$ (highly skewed).}
\label{fig:zipfquiv}
\end{figure}

The results for Zipf workloads in Figure \ref{fig:zipfquiv} look quite
different. At high skewness $s$, we find that Bloom filters are less useful and
it is better to allocate more memory to the buffer. At low skewness, the best
configuration is a mixture of mostly Bloom filter and cache memory with a
relatively small write buffer.

This effect may be due to the fact for highly skewed workloads, we obtain
better savings for the small hot set of keys by using the cache (for reads) and
the write buffer (for writes, and also as kind of auxiliary cache). For less skewed
workloads, we are more likely to request unpopular keys which may be buried
deep in the tree and impose a higher IO cost. 
To counteract this, we need
better Bloom filters.

\begin{figure}[!htb]
\begin{center}
\includegraphics[width=0.5\textwidth]{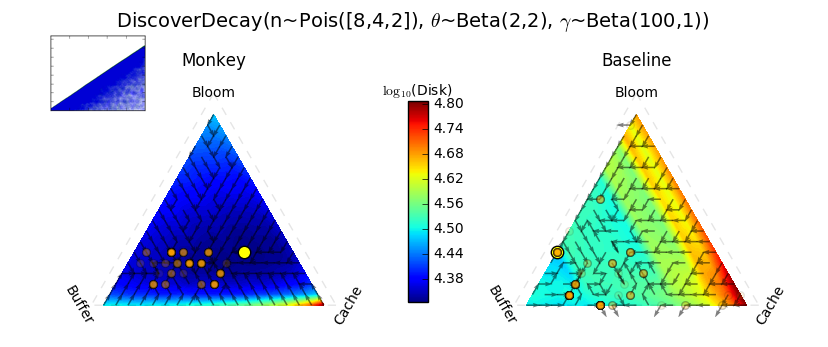}
\includegraphics[width=0.5\textwidth]{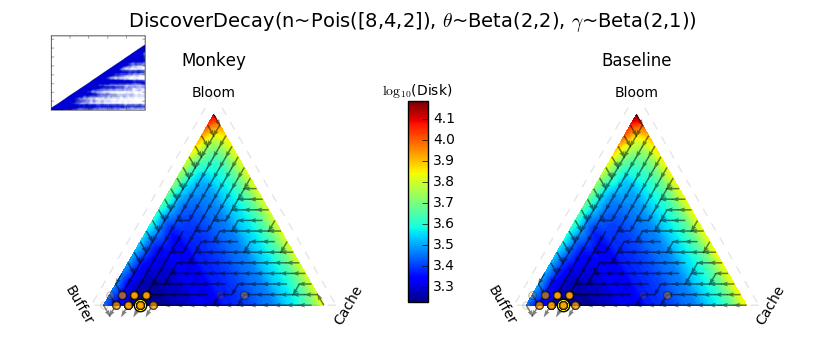}
\end{center}
\caption{Discover-Decay simulation results overlaid with gradient estimates for
    lightly skewed and highly skewed.}
    \label{fig:discdecquiv}
\end{figure}

\begin{figure}[!htb]
    \begin{center}
    \includegraphics[width=0.5\textwidth]{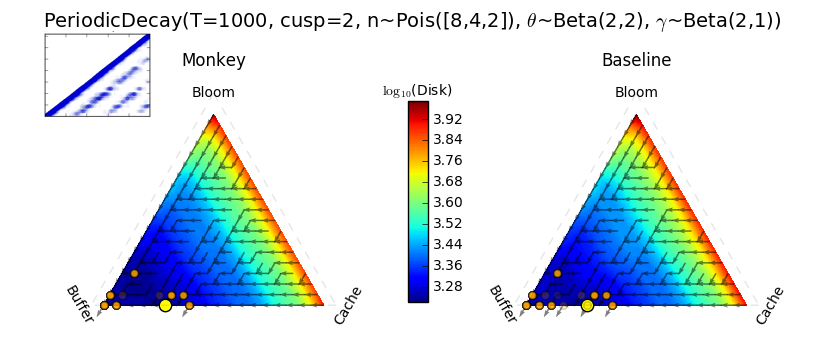}
    \includegraphics[width=0.5\textwidth]{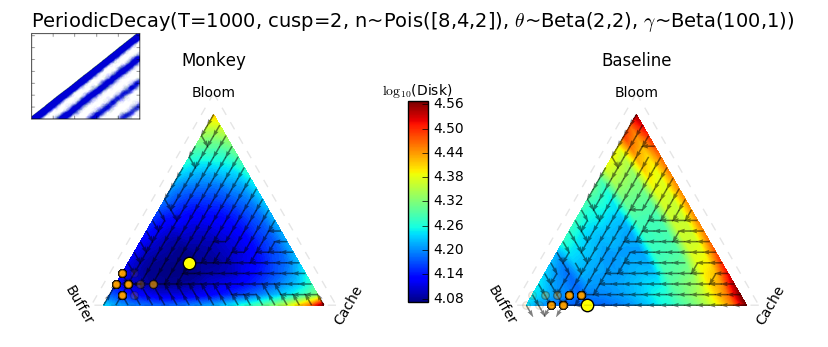}
    \includegraphics[width=0.5\textwidth]{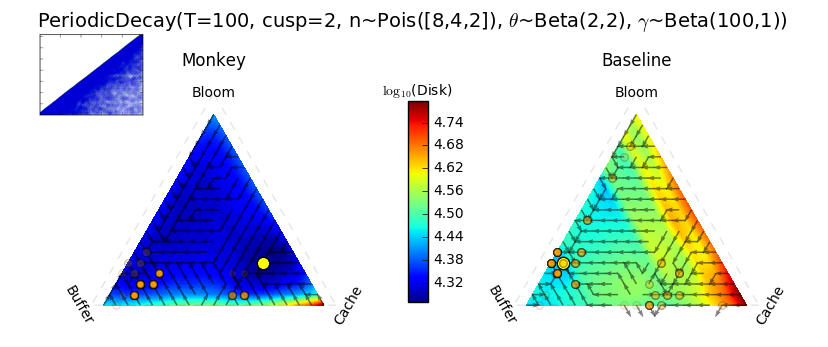}
    \end{center}
    \caption{Periodic Decay simulation results overlaid with gradient estimates for variations in periodicity (T) and skew (Beta distribution parameters).}
    \label{fig:periodquiv}
    \end{figure}

Finally, for the Discover-Decay and Periodic Decay workloads in Figures \ref{fig:discdecquiv} and \ref{fig:periodquiv}, we find that our gradients
capture the behavior we noted near the beginning of this Appendix section. For lower
effective numbers of popular keys (but high temporal locality), we tend
to end up allocating most of our memory to the buffer and none to the Bloom
filters, but as our ``working set'' expands, we are pushed closer to the center
of the graph. In the bottom row of Figure \ref{fig:periodquiv}, gradient descent is drawn into two distinct
modes based on the starting location, suggesting that our gradient estimations
are high-resolution enough to capture the nonconvexity. In general, there are
many situations in which increasing either the write buffer or the Bloom filter will
reduce I/Os, so we should expect multiple locally optimal allocation strategies
to exist.

\subsection{Modeling} \label{modeling}

We first consider the case of a uniform query distribution and then show how
the formulation can be generalized to any distribution with an empirical trace.

\subsection{Uniform query distribution}

\noindent Assuming we have
\begin{itemize}
\itemsep-1em
\item $N$ items in total DB \\
\item $E$ size of an entry in bits \\
\item $M$ total memory \\
\item $M_c$ memory allocated to cache \\
\item $M_b$ memory allocated to write buffer\\
\item $B$ entries that fit in a disk page \\
\item $P_b$ size of the write buffer in pages, i.e. $\frac{M_b}{BE}$ \\
\item $T$ ratio between levels of LSM-tree such that \\
\item $L1 = T * P_b * B$, $L2 =T^2 * P_b*B $, and so on,
\end{itemize}

\noindent then we can solve for $L$ the total number of levels required to store all the data: \\
$$P_b*B * \frac{1-T^L}{1-T} = N$$
$$L= \lceil \textrm{log}_{T} \left(\frac{N(T-1)}{BP_b} + 1\right) \rceil$$

The average cost of a write remains the same as for the basic LSM-tree case:

$$
\text{write cost} = \textrm{log}_{T} \frac{N}{P_bB}
$$

The average cost of a read must be considered probabilistically over all
possible locations of the read item, in this case assuming a uniformly random
distribution of reads:

\begin{itemize}
\item Probability that read is in write buffer $= p(\text{MT}) = \frac{P_b*B}{N}$
\item Probability that read is in cache $= p(\text{cache}) = \frac{M_c/E}{N}$
\item Probability that read is in L1 but not in cache $= p(L1)$ $$= \frac{P_b*B * T - \frac{P_b*B*T}{N-P_b*B} * M_c/E}{N}$$
\end{itemize}

where the numerator is the number of items $P_b*B*T$ that are in the first level
minus the proportion of items from that level that are probabilistically in the
cache already: $$\frac{P_b*B*T}{N-P_b*B} * M_c/E$$ and finally where the $N-P_b*B$
comes from the fact that items already in buffer (L0) are not allowed to
occupy the cache.

Therefore, given a uniform query distribution, the full expected cost in disk
reads of a read is
$$E[C_{\text{uniform}}] = p(\text{MT}) * 0  + p(\text{cache}) * 0 + \sum_{i=1}^L p(L_i) * i$$
$$=\sum_{i=1}^L \frac{P_b*B * T^i - \frac{P_b*B*T^i}{N-P_b*B} * M_c/E}{N} * i$$

\subsection{Bloom Filters}

The previous analysis hasn't yet accounted for the presence of Bloom filters,
which reduce the likelihood we will unnecessarily access a larger level. For a
Bloom filter of $k$ bits with $h$ independent hash functions $h_1, h_2,...h_h$,
the probability that a given bit is still set to 0 after inserting $n$ keys is 

$$
(1 - \frac{1}{k})^{n*h}
$$
Then the probability of a false positive is 
$$
(1- (1 - \frac{1}{k})^{n*h})^h \approx (1 - e^{-hn/k})^h
$$

We can minimize this over $h$ to find the optimal number of hash functions,
which is $h = \mathrm{ln}(2) * \frac{k}{n}$. Assuming that this is the number
of hash functions $h$ we will use, the probability of a false positive as a
function of the number of bits is then 

$$
(1 - e^{-\mathrm{ln}(2)*k/n*n/k})^{\mathrm{ln}(2) * \frac{k}{n}} = (\frac{1}{2}) ^ {\mathrm{ln}(2) * \frac{k}{n}} \approx (.6185) ^  {\frac{k}{n}}
$$

For an item in any level $L_i$ of the LSM-tree with $i \geq 2$ we can
reduce the expected cost of accessing that item from $L_i$ by the number of Bloom
filter negatives at any level $j<i$. \\ \\ Then the expected cost of accessing
an item at $L_i$ is  $$\sum_{j=1}^{i-1} p(fp_j) * 1 + 1$$ Where $p(fp_j)$ is
the probability of a false positive for that key at level $j$ and 1 is the cost
of actually accessing the item at level $i$ assuming fence pointers that lead
us to the correct page.

\subsection{Cost with Bloom Filters - Base Case}

Assuming a random distribution of reads, we now consider also the probability that a Bloom filter allows us to ignore a level: \\
Expected cost of read for an item in the tree = $$p(mt) * 0  + p(cache) + 0 + \sum_{i=1}^L p(Li) * \sum_{j=1}^{i-1} p(fp_j)$$ \\
Expected cost for a null result read = $\sum_{j=1}^{L} p(fp_j)$

Given a total memory allocation $M$, the total number of bits we can allocate
to Bloom filters is $M-M_c = \sum_{i=1}^L m_i$ \\ Then the total formula for
the expected cost of a read in the tree is: 

\begin{multline}
$$E[c] = \sum_{i=1}^{L} \frac{B*P*T^i - \frac{P_b*B*T^i}{N-P_b*B} * M_c/E}{N} \\ \cdot \left[ \left(\sum_{j=1}^{i-1} (.6185) ^  {\frac{m_j}{P_b*B*T^j}}\right) +1 \right]$$ 
\end{multline}

Whereas with a given percentage of null reads in the workload $p_{null}$:

\begin{multline}
$$E[c] = (1-p_{null})\sum_{i=1}^{L} \frac{P_b*B*T^i - \frac{P_b*B*T^i}{N-P_b*B} * M_c/E}{N} \\ \cdot \left[ \left(\sum_{j=1}^{i-1} (.6185) ^  {\frac{m_j}{P_b*B*T^j}}\right) +1 \right] + p_{null}\sum_{j=1}^{L} p(fp_j)$$
\end{multline}
\begin{multline}
$$E[c] = \sum_{i=1}^{L} (1-p_{null})\frac{P_b*B*T^i - \frac{P_b*B*T^i}{N-P_b*B} * M_c/E}{N} \\ \cdot \left[ \left(\sum_{j=1}^{i-1} (.6185) ^  {\frac{m_j}{P_b*B*T^j}}\right) +1 \right] + p_{null} \cdot p(fp_i) $$
\end{multline}

\subsection{Cost Gradients w. Bloom Filters - Generalized Distribution}

\subsubsection{Cache Gradient}

Note that in the above, the workload specific factors are the probability that
a read is at any given level and the related probability that any given item
from a level is already in the cache. To compute an empirical estimation of the
probability that any given item is in a level but not already in the cache, we
can simply keep statistics on the total number of times a key was found in that
level divided by the total number of (non-null) read queries executed. Then we
can consider the following simplification:

\begin{multline}
$$E[c] = \sum_{i=1}^{L} (1-p_{null})\left[p(L_i) - \frac{p(L_i)}{(N - P_bB)} * M_c/E \right]\\ \cdot \left[ \left(\sum_{j=1}^{i-1} (.6185) ^  {\frac{m_j}{P_b*B*T^j}}\right) +1 \right] + p_{null} \cdot p(fp_i) $$
\end{multline}

Taking the derivative with respect to the number of entries in the cache,
$M_c/E$, we get:

$$
\sum_{i=1}^{L}  -(1-p_{null}) p(L_i)/(N - P_bB) \cdot \left[ \left(\sum_{j=1}^{i-1} (.6185) ^  {\frac{m_j}{P_b*B*T^j}}\right) +1 \right]
$$

Which is just the average cost of a read throughout the tree. Then, to keep
statistics on how valuable we expect the cache to be, we maintain statistics on
the average cost of every read performed in the window of interest.

\subsubsection{Bloom Filter Gradients}

Because the memory allocation problem is discrete anyway, we consider the value
of the Bloom filters as a finite difference, that is the approximate value of
any marginal bloom filter bit at level $k$ will be $E[c | m_k+1] - E[c | m_k]$.
In this computation, all terms in the sums drop out except for those concerning
$m_j$, and we are left with:

\begin{multline}
$$\sum_{i=k}^{L} (1-p_{null})\left[p(L_i) - \frac{p(L_i)}{(N - P_bB)} * M_c/E \right] \\ \cdot \left\{ \left[ \left( (.6185) ^  {\frac{m_k+1}{P_b*B*T^j}}\right) +1 \right] - \left[ \left( (.6185) ^  {\frac{m_k}{P_b*B*T^j}}\right) +1 \right] \right\} \\+ p_{null}  \left( (.6185) ^  {\frac{m_k+1}{P_b*B*T^j}} - (.6185) ^  {\frac{m_k}{P_b*B*T^j}}\right)$$
\end{multline}

Rearranging terms, we get:
\begin{multline}
$$\sum_{i=k}^{L} \left[(1-p_{null})\left[p(L_i) - \frac{p(L_i)}{(N - P_bB)} * M_c/E \right] +  p_{null} \right] \\ \cdot \left( (.6185) ^  {\frac{m_k+1}{P_b*B*T^j}} - (.6185) ^  {\frac{m_k}{P_b*B*T^j}}\right)$$
\end{multline}

Where this is exactly the number of times the given bloom filter is accessed
times the difference in the theoretical false positive rates given memory
allocations $m_j$ and $m_j+1$. Then, to keep statistics on how valuable we
expect any given Bloom filter to be, we maintain statistics on the number of
times every Bloom filter was accessed in the window of interest.

\subsubsection{Write Buffer Gradient: Gets}

To estimate the additional value of any marginal memory in the write buffer with
respect to reads, we must make a number of simplifications, as $P_b$, the number
of pages in the write buffer, factors into every term in this equation. Further, the
interaction between $P_b$ and most of the terms is not available in closed form,
in general. Rather, the critical terms $P(L_i)$ we are empirically estimating.
Then, for reasonably large values of $N$ and $P_b$, we will assume that the Bloom
filter false positive rate stays approximately the same, as does the value of
the cache. Then, we consider only the change in I/Os occurring from the altered
probability of any given element occurring in any level as a result of more
elements being in the write buffer. We can provide a simple estimate
of this by assuming that any items we add to the write buffer would have
otherwise occurred in L1, and in the resulting cascade, $T^{i}$ times that number of items will be moved up into 
each level $L_{i}$ from the level below.

Then, an appropriate estimate of how useful any additional space of memory in
the write buffer is for reads is simply the resulting change in $p(L_i)$ for each level (that is, the number of hits
we expect to see on the newly added elements) $* fp_{i}$ for any level $i\neq0$, as the original cost of accessing that element was $\sum_{j=1}^i fp_{j} + 1$, and the new cost of accessing is $\sum_{j=1}^{i-1} fp_{j} $, the difference between which is just $fp_{i}$. For $i=0$, the write buffer itself, the expected savings per hits is exactly 1, as the item will be moved from having an access cost of 1 to 0. To estimate how many additional times L1 would be accessed if
we instead allocated the final portion of the write buffer to L1, we keep
statistics on how often the final spots of the write buffer were accessed in a
read. In practice, these spots are accessed only very infrequently, as the
write buffer is accessed only a handful of times at this stage before being flushed.
This statistic might be more helpful on a system with constant compaction
rather than a full level flush. For the rest of the levels, we simply assume the same hit rate per key
as measured over the existing keys on any level and multiply by the number of elements we will be adding to calculate
the expected accesses to the new keys on each level. We then multiply by the empirical rate of bloom filter false positives on the level.

\subsubsection{Write Buffer Gradient: Puts}

For the write buffer, we must additionally consider the saved I/Os for the update/insert operations.  $$
\text{write cost} = \textrm{log}_{T} \frac{N}{P_bB} $$ Taking the derivative with
respect to $P_bB$, the number of items in the buffer, we get $\frac{1}{P_bB}$ In
discrete terms, this evaluates to $\textrm{log}_{T} \frac{P_bB}{P_bB+1}$. 

Unfortunately, this simplification only works if we can assume that memory is being allocated in 
page-size chunks and that the workload has no duplicates. In practice, the number of I/Os associated
with reading and writing throughout the merging process is a stepwise function that depends on page size, as reading or
writing one element from or to a page has the same I/O cost as reading or writing a full page. To simplify our analysis
of the page size write savings, we consider only a ratio of $T=2$, and we begin by addressing the case wth no 
duplicates. 

With no duplicates, the final number of elements at any level of the tree is a
deterministic function of the number of elements inserted as well as the level
sizes. Then considering the empirical number of items inserted into the buffer
as well as the size of the original buffer, we can solve for the theoretical
final structure of an alternate LSM-tree that had a buffer of size $P_bB + 1$.

Additionally, given the number of elements on any given level, no duplicates,
and an original buffer size $P_bB+1$, we know the number of times each
$T^{i}*(P_bB+1)$-size chunk on each level will have been read and written given
the current fullness of the level. We can then multiply these numbers of known
chunk reads and writes by the ceiling of the size of those possible chunks
(which, with ratio $T=2$ will be $T^{i}*(P_bB+1)$ and $T^{i}*(P_bB+1)*2$) divided
by pagesize, $B$. This gives us a more realistic number in which additions of
less than a pagesize of memory are not helpful in I/O savings. 

Comparing the read and write costs of this theoretical tree to the empirical
reads and writes accesses of the existing tree gives us an expected I/O savings
related to updates for the larger tree.

We consider additionally the fact that I/O savings are in general lessened by
the number of duplicates inserted, as duplicates will not be merged across the full
depth of the tree. To take this into account we also keep a statistic for the
total number of duplicates merged over the window of interest per level and use
this to calculate the percentage of duplicates removed relative to total keys
at each level. This factors in in several places. First, when computing the
theoretical allocation of keys in the final tree, we consider the total number
of items that would have come in to the buffer from the empirical
count and adjust this at each level by the percentage that are expected to have
been removed as duplicates. Further, when computing read and write I/Os during
merging, we expect that number of items written when the level is already half
full should be decreased by the expected number of duplicates removed among the
two sets of keys. Again, the resulting I/O savings will be stepwise in
pagesize. In particular, if the original size of the array would have only been
slightly into the final page, it will take very few duplicates to reduce the
I/O count by 1, whereas if all pages would have been full, it will take a full
page's worth of duplicate removals to improve I/Os. The same savings will be
experienced again when these items are read to be merged into the lower level.

The correct way to handle the duplicates requires somewhat more consideration,
but the only statistics we are currently using The are the empirical number of
update queries and the empirical number of duplicates found and removed on each
level over the window of interest.

\subsection{Estimating Statistics with $O(1)$ Memory}

\textbf{Cache:} to estimate the number of storage accesses we will save by adding
$dM$ extra bits of memory to the cache, we let consider $dM$ as a number of
extra entries in the cache. That is, we calculate the savings from having
$dM/E$ extra cache entries available. As mentioned above, the relevant
statistic here is the average cost of a read in the database. To calculate
this, we collect statistics on the total number of storage accesses and total
number of queries. The expected cost per query is then the number of disk
accesses over the window divided by the total number of queries. To approximate
the probability of the item being in the cache times the number of queries, we
maintain a statistic for the number of times the last cache slot was accessed
during the window of interest and make the assumption that the number of hits
on the next marginal slot(s) would be approximately the same. Then we can
calculate the final expected I/O savings as

$$dM/E * E[hits] * E[cost/query]$$

\textbf{Bloom Filters:} To estimate the number of storage accesses we will save by adding $dM$ extra bits of memory to the Bloom filters, we first decide how to allocate that $M'_{bloom} = M_{bloom}+dM$ bits using Monkey or the baseline allocation, giving us $m_i$ and $m'_i$ bits per Bloom filter on each level. At each level $i$, for both $m_i$ and $m'_i$, we update rolling averages of the theoretical false positive rate $\hat{fp}_i = \mathbb{E}\left[0.6185^{\frac{m_i}{n_i}}\right]$ and $\hat{fp}'_i = \mathbb{E}\left[0.6185^{\frac{m'_i}{n_i}}\right]$ every time the Bloom filter is queried (where $n_i$ is constantly changing based on insertions and flushes of the filter). These statistics (individual floats) give us an estimate of the aggregate false positive rate at $m_i$ and $m'_i$ robust to changing level fullness. Finally, we keep a counter $n_{i,\text{bloom false}}$ of the number of times requested items are \textit{not} in bloom filter $i$. This counter is incremented either when the bloom filter returns false (which we know immediately) or returns a false positive (which we can record after fruitlessly searching the level). This counter allows us to estimate storage accesses resulting from our current or altered false positive rates. The final savings is therefore \[
  \text{Savings}(M'_{bloom}) = \sum_{i} (\hat{fp}'_i - \hat{fp}_i) * n_{i,\text{bloom false}},
\]

and only requires keeping two floats and one integer. Note that in our simulation, for flexibility, we keep a histogram of $n_i$ values at each bloom filter request to avoid needing to predetermine $m'_i$, but in a practical implementation this is unnecessary.

\begin{figure}[!htb]
\begin{center}
\includegraphics[width=0.33\textwidth]{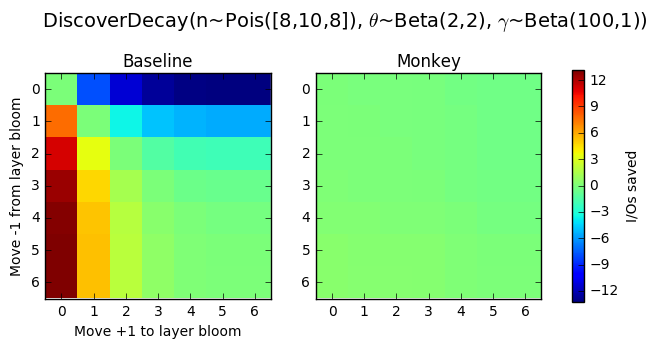}
\includegraphics[width=0.33\textwidth]{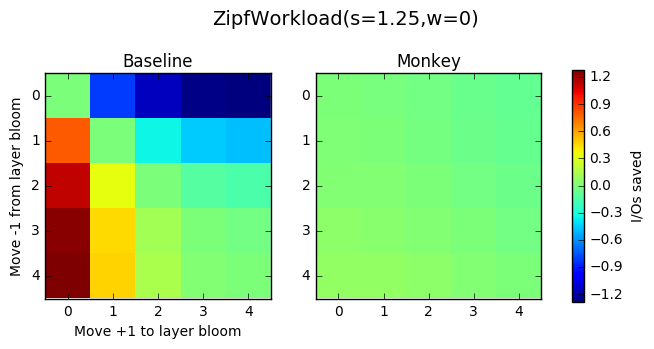}
\end{center}
\caption{Estimated change in I/Os when moving bits from one Bloom filter to another (keeping total bloom filter memory constant). Regardless of workload, changes in I/Os for Monkey are all less than 1, indicating its optimality.}
\label{fig:bloom-realloc}
\end{figure}

Note that because we can obtain these estimates on a level-by-level basis, we can investigate whether reallocating memory from one Bloom filter to another, empirically, should reduce I/Os. Validating the results of Monkey \cite{Dayan2017}, in Figure \ref{fig:bloom-realloc} we find that for the baseline allocation, moving bits does improve performance, but for Monkey, it does not, regardless of workload.

\textbf{Buffer:} To estimate the number of storage accesses we will save in reads by adding $dM$ extra bits of memory to the buffer, we use statistics maintained on the total bloom filter accesses per level, Bloom filter false positives per level, and hits per level. We estimate the expected
additional number of hits on any given level as the original hits times the new theoretical size divided by the actual original size.
That is, the number of extra hits is equal to $$new\_hits_{i} = hits_{i} * \frac{size_{i} + dM*T^{i} }{ size_{i}}$$
For each expected hit, we have an I/O savings equal to the false positive rate on the bloom filter of that level, as described in the previous section. To calculate this for a level $i$, we use 
$$E[savings/hit]_{I} = \frac{false\_positives_{i}}{bloom\_accesses_{i}}$$
Then the total number of I/Os saved should be 
$$
\sum_{i=0}^L new\_hits_{i} * E[savings/hit]_{i} 
$$
where for level 0, the write buffer, the $E[savings/hit] = 1$, as the access cost at L1 is always exactly 1 and the access cost at 
the write buffer is always 0.

To estimate the number of storage accesses we will save in writes/updates by adding $dM$ extra bits of memory to the write buffer, 
we maintain statistics on total number of entries that passed through any given level, number of duplicates removed at any given level, and number of entries in any given level at the end of the period. For a workload without duplicates, we can simply
use these statistics to deterministically calculate the final allocation and number of read and write I/Os that would have
occurred throughout the process for a second tree with write buffer size + $dM$, calculating every batch of read and write merges and summing over the number of pages that would have been involved. For the original tree we can either use statistics on empirical I/Os during the merging process or use the same deterministic formula to calculate what they would have been. The expected saved I/Os then is simply

$$cost_{tree} - cost_{tree+dM}$$

When we consider duplicates, the estimate becomes much more noisy. To consider the effect of duplicates on reducing the total number of pages read and written during the merging process, we reduce the number of entries that pass through each level of our theoretical larger tree by the percentage of duplicates removed at each level, calculated as $$\frac{duplicates\_removed_{i}}{total\_entries_{i}}$$

This then changes the final level structure of the estimated tree. We also consider that duplicates should reduce the total number of entries written and then read after two segments are merged together. Then for those read and write components that occur on an already half-filled level, we reduce the number of elements by multiplying by $$1 - \frac{duplicates\_removed_{i}}{total\_entries_{i}}$$
This will reduce the total I/Os by number of page reads it makes unnecessary. With this adjusted cost for the larger tree, we again calculate the expected saved I/Os as the estimated I/Os of the hypothetical larger tree subtracted from the empirical or theoretical I/Os of the existing tree.

\subsection{Gradient Validation}

\begin{figure*}[!ht]
\centering
\includegraphics[width=0.8\textwidth]{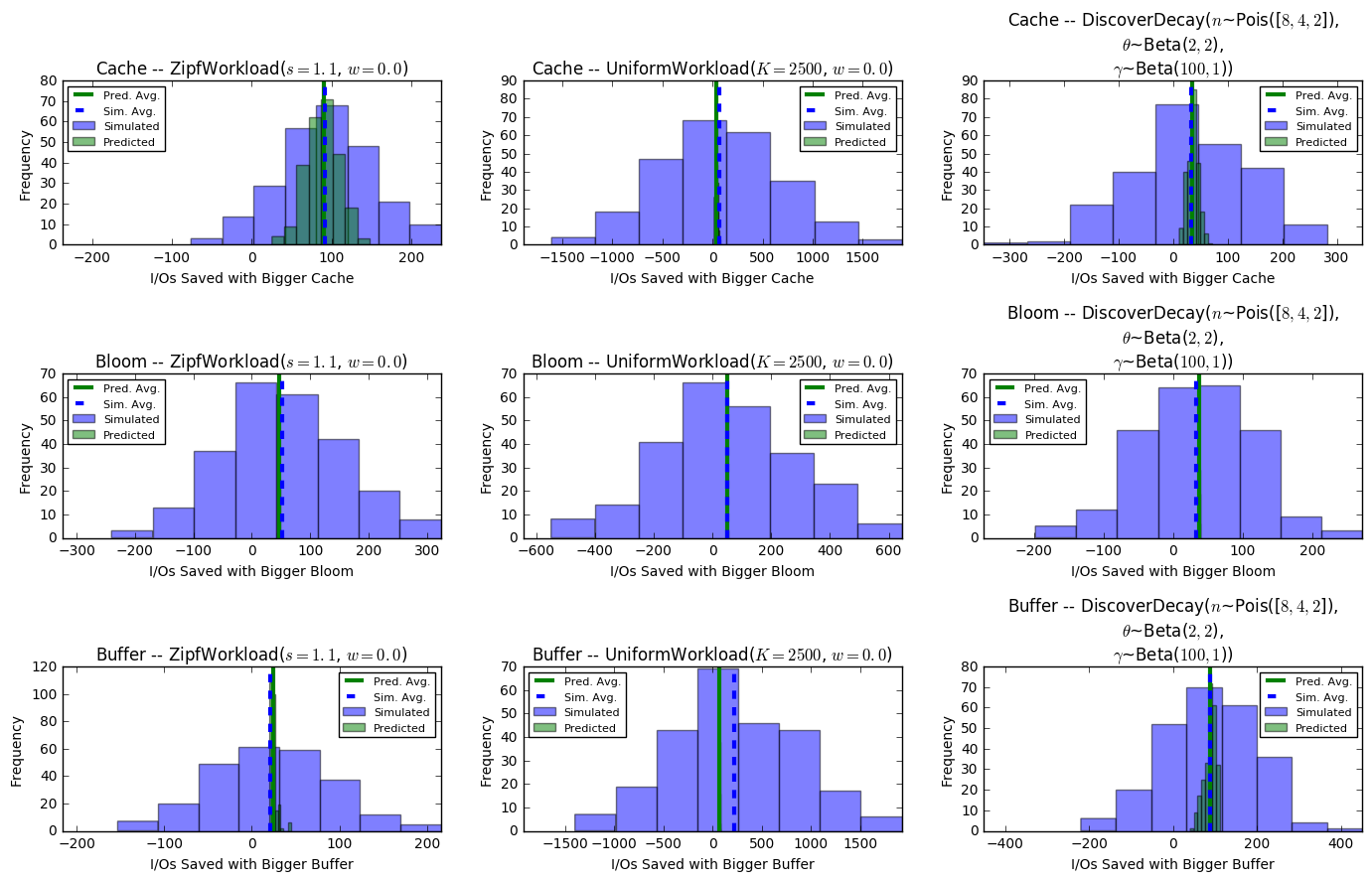}
\caption{Light-footprint statistical estimations of the gradient vs. simulated
results for cache, Bloom filters, and the write buffer on three distinct workloads.}
\label{fig:savings}
\end{figure*}

To confirm that our estimates are reasonable, we ran 250 simulations for three
separate workloads and compared our estimates of each gradient to the actual
savings for a separate tree with 8 bytes of extra memory in the corresponding
LSM component (against which we ran the same workload). Results can be seen in
Figure \ref{fig:savings}.

There is a large amount of variance in the simulated results, both because of
randomness in the separate instantiations of the workload and randomness in
the execution of its queries, but for the most part, our estimates of the average
savings are both precise and accurate. There is a slight deviation for the uniform
buffer savings calculation, but the variance is so high that it does not appear
to be significant.

The fact that our estimates of the expected I/O savings are so precise across
workloads gives us confidence first that our simulation and modeling are correct,
and second that they will generalize to more complex, real-world workloads
with more queries and keys.

\end{document}